\documentclass[sigconf,10pt,nonacm]{acmart}

\AtBeginDocument{%
  }

\usepackage{amsmath,amssymb,amsfonts}
\usepackage{algorithmic}
\usepackage{graphicx}
\usepackage{textcomp}
\usepackage{xcolor}
\usepackage{xspace}

\usepackage{enumerate}
\usepackage{url}
\usepackage{booktabs}
\usepackage{graphicx}
\usepackage{subfigure} 
\usepackage{bm}
\usepackage{multirow}
\usepackage{array}
\usepackage{makecell}
\usepackage{tabularx}
\usepackage{stfloats}
\usepackage{makecell}
\usepackage{pifont}
\usepackage{enumitem}

\renewcommand\footnotemark{} 
\begin{document}


\title{Ghost Points Matter: Far-Range Vehicle Detection with a Single mmWave Radar in Tunnel}

\renewcommand{\shorttitle}{Ghost Points Matter: Far-Range Vehicle Detection with a Single mmWave Radar in Tunnel}

\author{
  Chenming He$^\dag$, Rui Xia$^\dag$, Chengzhen Meng$^\dag$, Xiaoran Fan$^\ddag$, Dequan Wang$^\dag$, Haojie Ren$^\dag$,  Jianmin Ji$^\dag$, Yanyong Zhang\textsuperscript{$\dag \; \diamondsuit$ *}\\
  \small $^\dag$University of Science and Technology of China, $^\ddag$Independent Researcher,\\
  $^\diamondsuit$Institute of Artificial Intelligence, Hefei Comprehensive National Science Center\\
  \{hechenming, xrxx, czmeng\}@mail.ustc.edu.cn, 
  gunanjiluzhe@gmail.com, 
  \{wdq15588, rhj\}@mail.ustc.edu.cn, \\
  \vspace{-5pt}
  \{jianmin, yanyongz\}@ustc.edu.cn
  }
\authornote{Corresponding author}

\renewcommand{\shortauthors}{}

\begin{abstract}
  Vehicle detection in tunnels is crucial for traffic monitoring and accident response, yet remains underexplored. In this paper, we develop mmTunnel, a millimeter-wave radar system that achieves far-range vehicle detection in tunnels. The main challenge here is coping with ghost points caused by multi-path reflections, which lead to severe localization errors and false alarms. Instead of merely removing ghost points, we propose correcting them to true vehicle positions by recovering their signal reflection paths, thus reserving more data points and improving detection performance, even in occlusion scenarios. However, recovering complex 3D reflection paths from limited 2D radar points is highly challenging. To address this problem, we develop a multi-path ray tracing algorithm that leverages the ground plane constraint and identifies the most probable reflection path based on signal path loss and spatial distance. We also introduce a curve-to-plane segmentation method to simplify tunnel surface modeling such that we can significantly reduce the computational delay and achieve real-time processing.

  We have evaluated mmTunnel with comprehensive experiments. In two test tunnels, we conducted controlled experiments in various scenarios with cars and trucks. Our system achieves an average F1 score of 93.7\% for vehicle detection while maintaining real-time processing. Even in the challenging occlusion scenarios, the F1 score remains above 91\%. Moreover, we collected extensive data from a public tunnel with heavy traffic at times and show our method could achieve an F1 score of 91.5\% in real-world traffic conditions.
\end{abstract}




\maketitle

\section{INTRODUCTION}
\label{sec:intro}

Nowadays, vehicle detection with roadside sensors plays a vital role in intelligent transportation systems, which enables real-time monitoring of traffic flows and supports accident response. Among various road types, tunnels represent a critical yet underexplored environment. The enclosed nature of tunnels heightens the risk of severe crash injuries and complicates rescue operations~\cite{tunnel_safety}. Therefore, achieving accurate vehicle detection in tunnels is an urgent task that demands immediate attention. 

Some researchers have developed effective vehicle detection systems using roadside cameras or LiDARs~\cite{bevdepth, bevheight, vips, vrf, trajmatch, UniSense}, but these systems still have limitations. Camera-based systems, in particular, struggle to provide accurate depth information for detected targets~\cite{gard} and are highly sensitive to lighting conditions~\cite{light-condition}. Meanwhile, LiDAR-based systems are expensive and have a limited sensing range. For instance, an 80-line LiDAR (RS-Ruby Lite~\cite{RS-Ruby}) costs over 15,000 dollars, yet its effective sensing range is limited to 120 meters. Moreover, the existing works only focus on open roads and lack research in tunnel environments. In response to these limitations, this study explores the use of millimeter-wave (mmWave) radar to develop a low-cost roadside system capable of far-range vehicle detection in tunnels, with a detection range exceeding 300 meters.


\begin{figure*}
    \centering
    \setlength{\abovecaptionskip}{1pt}
    \includegraphics[width=0.97\linewidth]{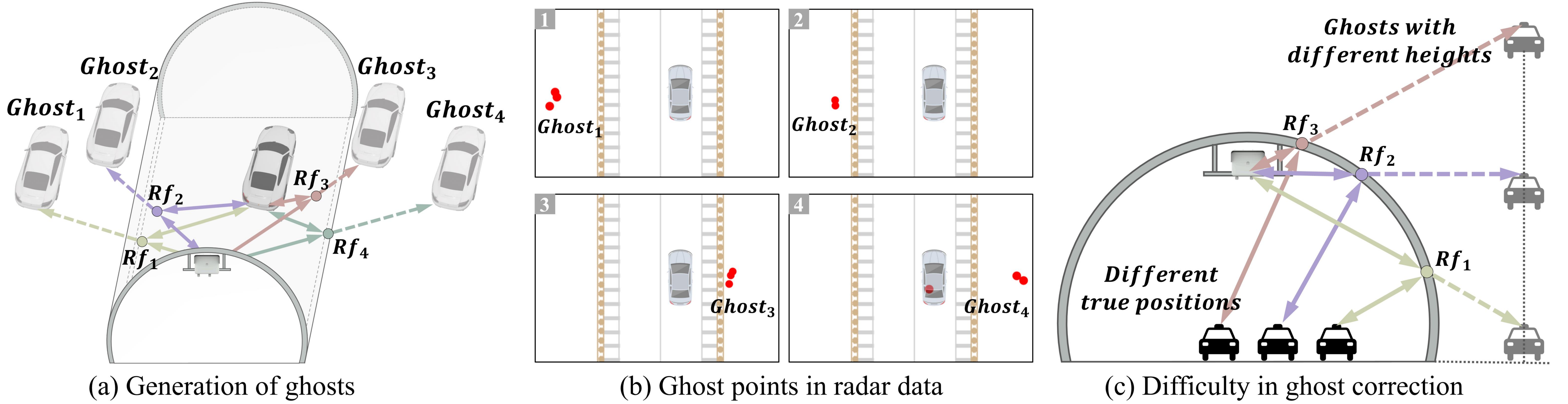}
    \caption{Ghost phenomena in tunnels and difficulty in ghost correction. \textmd{(a) Radar signals reflected from a vehicle may follow numerous 3D reflection paths. (b) Ghost points of the same vehicle appearing at different positions in nearby frames. (c) The absence of height information results in multiple possible reflection paths for a detected ghost.}}
    \label{fig:multi-ghost}
\vspace{-2mm}
\end{figure*}

The mmWave radar, due to its high ranging accuracy, long detection range, and all-weather reliability~\cite{k-radar}, has gained increasing attention in the research community for detection~\cite{ipsn1,M4esh,mmplace,grfall,mobicom-1,mobicom-2, Lt-fall,zhang2024single}. However, radar also has limitations. As radar signals propagate, they reflect off surfaces like the ground and walls, altering their propagation paths. This phenomenon is known as the multi-path effect~\cite{nsdi-multi,multipath1,multi-doppler, multipath}.
The reflected signals generate ghost points, which appear outside the roadway (as illustrated in Figure~\ref{fig:multi-ghost}), potentially leading to localization errors and false alarms. In previous studies, ghost points were traditionally regarded as interference~\cite{Griebel-pointnet,ghost-idf, mmOVD}. One intuitive and practical solution is to define the drivable region and filter out ghost points outside it, but this exacerbates the sparse nature of radar points, leaving insufficient data for accurate detection. 

In this work, we propose a novel perspective: \emph{\textbf{ghost points matter for better detection.}} Our key idea is that ghost points are not randomly generated. Instead, they result from multiple reflections of radar signals involving vehicles and tunnel surfaces, carrying critical clues about the true positions of vehicles. By modeling the tunnel and tracing the specular reflection process of radar signals between vehicles and tunnel surfaces, we can reconstruct the reflection paths of ghost points. This allows the ghost points to be corrected to their true positions, thereby improving vehicle detection. 

\begin{figure}
    \centering
    \setlength{\abovecaptionskip}{0mm}
    \includegraphics[width=0.95\linewidth]{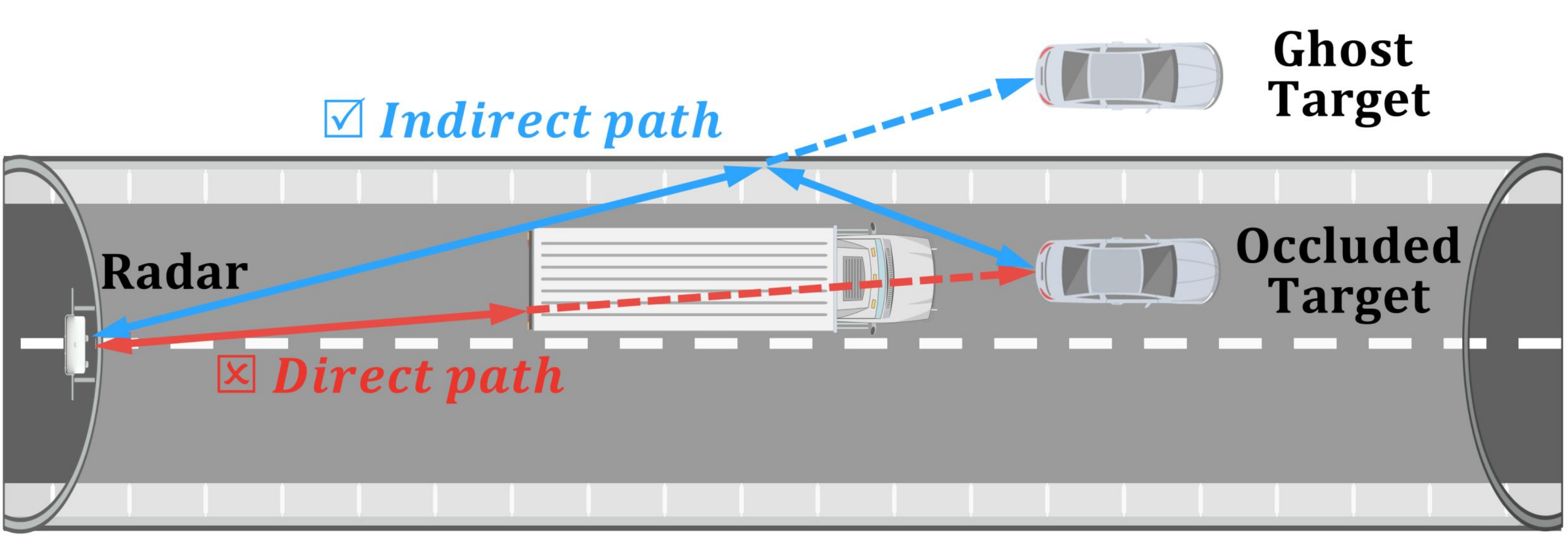}
    \caption{Indirect reflection paths of signals create ghost targets, which can be used to localize the fully occluded vehicles.} 
     \label{fig:ghost}
\vspace{-2mm}
\end{figure}

Additionally, we observe that the limited ceiling height of the tunnel reduces the allowable sensor installation height to approximately 5 meters, compared to previous roadside systems typically installed on poles 7-10 meters high~\cite{Dair-v2x}. This lower installation height increases the likelihood of vehicle occlusion. Nonetheless, occluded vehicles can still generate ghost points, as the radar signals can bypass the blocking vehicle via indirect reflection paths, as shown in Figure~\ref{fig:ghost}. By leveraging these ghost points, we can accurately localize vehicles in tunnels, even when they are fully occluded. 

However, significant challenges arise due to the inherent properties of the tunnel environment and radar limitations. (1) \emph{Complex 3D reflections}. Unlike the previous works~\cite{mmOVD, street-corner} that only consider 2D plane reflections between radar signals and the ground or vertical walls, the enclosed and narrow tunnel environments lead to more complex 3D reflections. (2) \emph{~Limited 2D radar points}. Commercial traffic mmWave radars typically generate only 2D points and lack height information.
As illustrated in Figure~\ref{fig:multi-ghost}(a), radar signals reflected from a vehicle can follow numerous 3D reflection paths, resulting in ghost targets appearing at various positions. This analysis is further supported by experimental results: Figure~\ref{fig:multi-ghost}(b) depicts radar points (shown in red) from nearby frames, indicating that ghost points of the same vehicle appear at different positions outside the driving lanes, and the radar points provide only top-view coordinates without heights.
These limitations lead to two main challenges. 
First, the limited 2D radar data makes it infeasible to directly determine the 3D reflection paths from numerous possibilities due to the absence of height-related constraints. As illustrated in Figure~\ref{fig:multi-ghost}(c), the detected ghost points can correspond to multiple possible heights, each associated with different reflection paths and true positions, making the accurate correction of ghost points highly challenging.
Second, the curved tunnel surfaces significantly increase computational complexity when tracing reflection paths, requiring the solution of equations that cannot be solved analytically (detailed in Section~\ref{tunnel-modeling}). The use of numerical methods introduces significant delays, making real-time processing impractical.

This paper presents mmTunnel, a roadside radar system designed to achieve far-range vehicle detection in tunnels, even for occluded vehicles. Our core idea is to correct ghost points, which appear outside the lane due to wall reflections, to the true positions by recovering their signal reflection paths. Specifically, we first propose a curve-to-plane segmentation method for modeling the 3D curved surface of a tunnel. By approximating the tunnel surface as a series of adjacent planes, we simplify the complex curved surface reflection problem into manageable planar reflection scenarios. To balance modeling accuracy and computing efficiency, we optimize the segmentation parameters by ensuring that the localization error introduced by this approximation remains within the radar's resolution limit. This approach ensures detection accuracy while enabling real-time processing. 


Further, to recover signal reflection paths without height-related constraints, we propose a \emph{weakly constrained generation, multi-criteria selection} framework, inspired by the hypothesize-and-verify paradigm in computer vision~\cite{hypothesis}. First, by leveraging the prior knowledge that vehicles must drive on the ground plane as a weak constraint, we reconstruct 3D reflection paths from 2D radar data through geometric ray tracing. This generates a set of candidate reflection paths for each ghost point. Second, we identify the most probable reflection path by minimizing signal path loss and spatial distance. The selected path enables accurate correction of ghost points to their true positions, significantly enhancing vehicle detection. Finally, we perform point clustering and target tracking using the corrected radar points, achieving real-time detection of all vehicles in the tunnel.

\begin{figure*}
    \centering
    \setlength{\abovecaptionskip}{1mm}
    \includegraphics[width=0.99\linewidth]{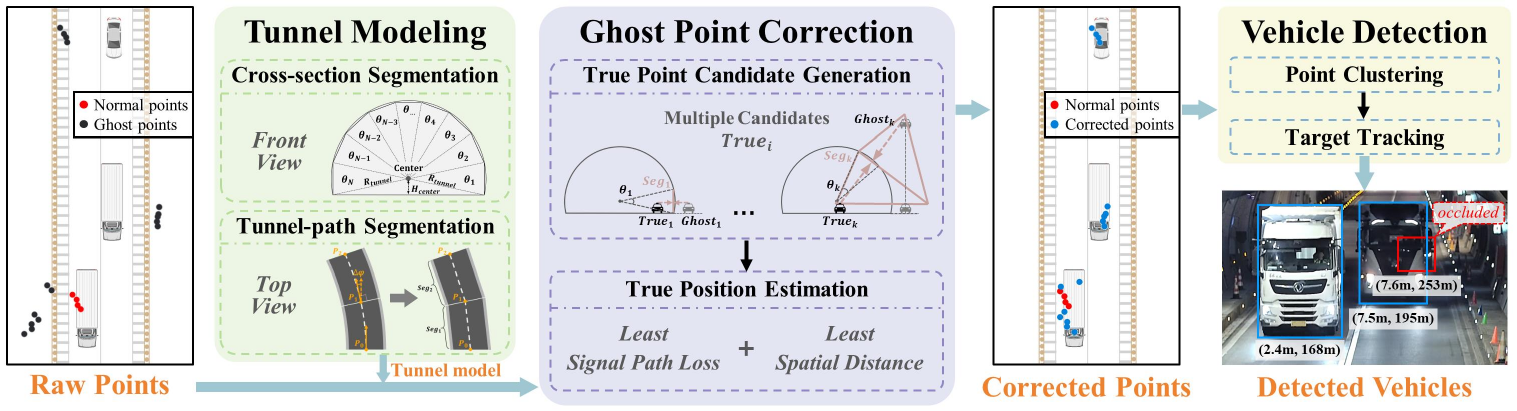}
    \caption{The mmTunnel system involves three main steps: (1) tunnel modeling, (2) ghost point correction, and (3) vehicle detection.} 
     \label{fig:mmTunnel}
\end{figure*}

In summary, we make the following contributions:

$\bullet$ We propose mmTunnel, a far-range vehicle detection system with a single traffic mmWave radar in the tunnel. To the best of our knowledge, this is the first roadside system capable of accurately detecting vehicles in tunnels, even when the vehicles are fully occluded.

$\bullet$ We devise a curve-to-plane segmentation-based tunnel modeling method that simplifies the complex 3D curved surface reflection problem into manageable planar reflection scenarios. 
We also design a two-stage ghost correction method that leverages the ground plane constraint and recovers the most probable signal reflection paths, thereby accurately correcting ghost points to their true positions.

$\bullet$ We extensively evaluated mmTunnel in both controlled scenarios and real-world traffic scenarios. In a straight tunnel and a curved tunnel, we conducted controlled experiments across various scenarios with cars and trucks. Our system achieves an average F1 score of 93.7\% for vehicle detection with 10 FPS. Even in the challenging congestion scenario (inter-vehicle spacing $< 2$ m) and occlusion scenario (a car fully occluded by a truck), the F1 score remains above 91\%. Moreover, we collected over 36,000 frames of data from a public tunnel, with up to 12 vehicles simultaneously within the radar's sensing range. Our system achieves an F1 score of 91.5\%, showing its robustness in complex real-world traffic.


\section{MMTUNNEL DESIGN}

\subsection{Problem Definition}

In this work, our objective is to continuously obtain the location of each vehicle in the tunnel. Formally, we are given the point cloud $\mathcal{P}$ from the traffic radar, which contains the top-view coordinates and Doppler velocity of radar points: $\mathcal{P}=\{p\},\; p=(x, y, v_d)$. The output of our system is the locations of all detected vehicles: $\mathcal{V}=\{V\},\; V=(x, y)$. As a common assumption in autonomous driving, we assume the radar pose parameters are known after the initial installation. Thus, all coordinates are unified into tunnel coordinate system, with the tunnel entrance as the origin.

For ease of explanation, this paper assumes the tunnel cross-section to be a circular segment, which is one of the most common shapes in real-world tunnels. However, our method can be easily adapted to tunnels with other cross-sectional shapes. The tunnel centerline, being a smooth curve, can be approximated using a polynomial function:
\begin{equation}
    \small y=a_0 + a_1 x + a_2 x^2 + \dots + a_nx^n.
\end{equation}
In practice, we found that cubic polynomials ($n=3$) suffice to accurately fit centerlines, as tunnels are typically designed with smooth and gentle curves to ensure driving safety.
Before subsequent processing, we label radar points located outside the predefined drivable lane boundaries as \emph{ghost points}, and those within the lane boundaries as \emph{normal points}. The lane boundaries are obtained from the tunnel construction design parameters and radar installation geometry. Our method then focuses on correcting ghost points to their true positions to improve vehicle detection accuracy.

\vspace{-3pt}
\subsection{System Overview}

Figure~\ref{fig:mmTunnel} depicts the overview of our mmTunnel system, which consists of three main steps. The tunnel modeling module segments the curved surface of the tunnel into a series of planes spliced together, thus providing a simplified yet effective model for the tunnel. Then, the ghost point correction module restores each ghost point to the most probable true position of the corresponding vehicle. Finally, the vehicle detection module uses the corrected points for point clustering and target tracking, ultimately generating the positions of all detected vehicles.

\vspace{-3pt}
\subsection{Tunnel Modeling through Curve-to-Plane Segmentation} \label{tunnel-modeling}

Our first consideration is modeling the tunnel surface, as it is essential for recovering the signal reflection paths and correcting the ghost points. The most intuitive solution is to measure the tunnel's cross-section and centerline parameters, which allows for modeling the curved surface. However, finding the reflection point on the curved surface for each ghost point requires solving complex equations, which are not analytically solvable (see Appendix~\ref{apd-curve}). Using numerical methods like Newton's method leads to significant delays. In our tests, this approach achieved only 6.2 FPS (frames per second), far below the radar's 10 FPS frame rate, making real-time processing infeasible. To solve this issue, we propose a curve-to-plane segmentation method for tunnel modeling, which simplifies the complex curved surface reflection problem into manageable planar reflection scenarios.

\begin{figure}
    \centering
    \setlength{\abovecaptionskip}{0mm}
    \includegraphics[width=0.98\linewidth]{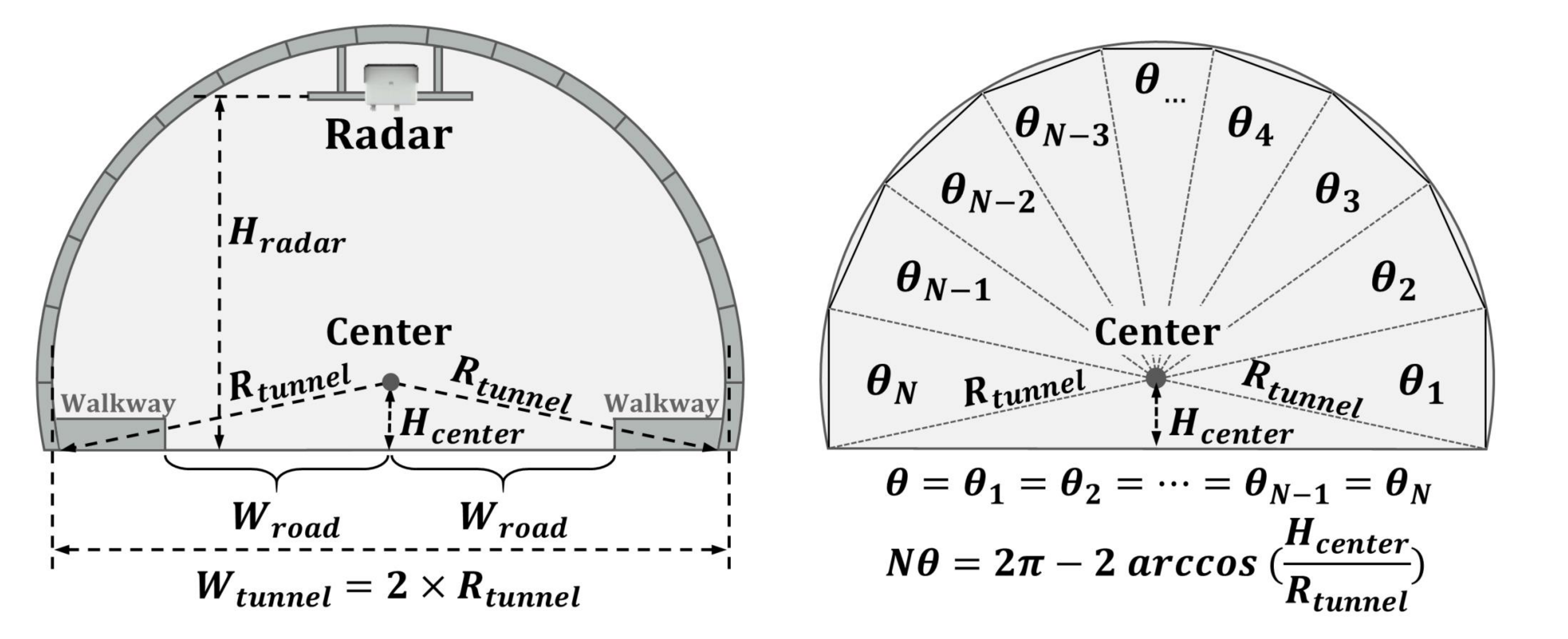}
    \caption{Cross-section segmentation in the front view. \textmd{The arc-shaped roof is divided into $N$ straight line segments.} }
     \label{fig:segment1}
\vspace{-3mm}
\end{figure}

\subsubsection{Cross-section Segmentation.} As shown in Figure~\ref{fig:segment1}, we first divide the circular arc-shaped roof of the tunnel into $N$ equal sectors and approximate each arc segment with a straight line. The angle of each sector is denoted as $\theta$. By measuring the circular radius ($R_{tunnel}$) and the height of the circular center ($H_{center}$) during the implementation, we can calculate the angle $\theta$ as follows:
\vspace{-3pt}
\begin{equation}
    \small \theta = \frac{1}{N} \left[ 2\pi - 2 \, \arccos \left( \frac{H_{center}}{R_{tunnel}} \right) \right].
    \label{eq:theta}
\end{equation}

\subsubsection{Tunnel-path Segmentation.} 
\begin{figure}
    \centering
    \setlength{\abovecaptionskip}{0mm}
    \includegraphics[width=0.98\linewidth]{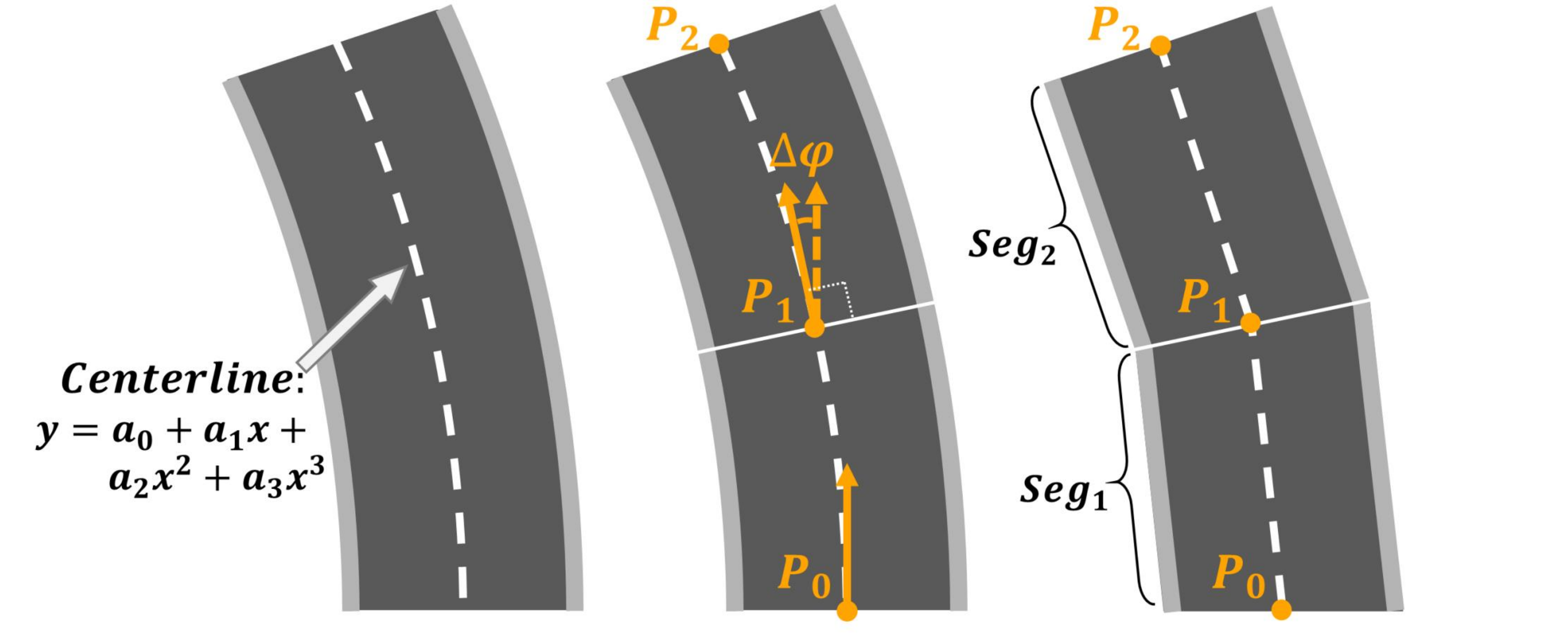}
    \caption{Tunnel-path segmentation in the top view. \textmd{The curved path is divided into $M$ straight path segments.} }
     \label{fig:segment2}
\vspace{-3mm}
\end{figure}

As shown in Figure~\ref{fig:segment2}, we iteratively divide the curved tunnel path into multiple straight segments. The dividing points of segment \( Seg_i \) along the centerline are denoted as \( P_{i-1} \) and \( P_i \). To control the error between the straight segments and the original curved path, the angle between the tangent lines at two adjacent dividing points on the centerline (denoted as \( TL_{i-1} \) and \( TL_{i} \)) is constrained within a preset threshold \( \Delta\varphi \):
\vspace{-3pt}
\begin{equation}
    \small \angle \left( TL_{i-1}, TL_{i} \right) = \Delta\varphi, \quad i = 1, \dots, M.
\end{equation}
Taking the plane perpendicular to $TL_{i}$ as the dividing interface and the line segment $\overrightarrow{P_{i-1}P_i}$ as the new centerline, we have the $i$-th straight path segment. At the start of this iteration, the first dividing point $P_0$ is the point on the centerline where $y = 0$. Using this method, we iteratively calculate the dividing points and obtain $M$ straight path segments.

\subsubsection{Parameter Optimization.} With the above segmentation methods, the curved tunnel surface is divided into $N \cdot M$ plane segments. Two key parameters govern this segmentation: sector angle $\theta$ and threshold angle $\Delta\varphi$. The smaller these two angles are, the more precise the model becomes. However, reducing these angles increases the values of $N$ and $M$, which in turn raises the computational load for the ghost point correction module, leading to system delays. To address this trade-off, we conduct a theoretical analysis to calculate the maximum localization error introduced by segmentation. We then determine the largest permissible values of $\theta$ and $\Delta\varphi$ that ensure the error remains within the radar’s resolution limit, thereby optimizing the segmentation parameters. Due to space limitations, we provide the detailed derivation in Appendix~\ref{apd-deriv}. The final results are as follows: 
\vspace{-3pt}
\begin{equation}
    \small E_{c} = 2R_{tunnel} \left( \sin\frac{\theta}{2} + \sin^2\frac{\theta}{2} \right), \enspace E_{p} = L_{max} \tan\Delta\varphi,
    \label{eq:error}
\end{equation}
where $E_c$ and $E_p$ represent the maximum errors from cross-section segmentation and tunnel-path segmentation, respectively. We also find that the lengths of straight path segments affect the localization error. To mitigate this, we limit the length of each path segment to a maximum threshold $L_{max}$. In our implementation, we set $L_{max} = 100$ meters. With the radar's range resolution of $2$ meters and the tunnel radius of $5.5$ meters, solving Equation~\ref{eq:error} gives the following parameter limits: $\theta < 18.08^\circ$, $\Delta\varphi < 1.15^\circ$. We applied the modeling algorithm in a curved tunnel (detailed in Section~\ref{sec:evaluation}). The tunnel roof is divided into 12 equal segments, and the tunnel path is divided into 7 segments, with each path segment ranging from 14 meters to 78 meters in length.

\vspace{-5pt}
\subsection{Ghost Point Correction through Two-stage Ray Tracing}\label{sec:ghost-correct}

\begin{figure}
    \centering
    \setlength{\abovecaptionskip}{0.5mm}
    \includegraphics[width=0.95\linewidth]{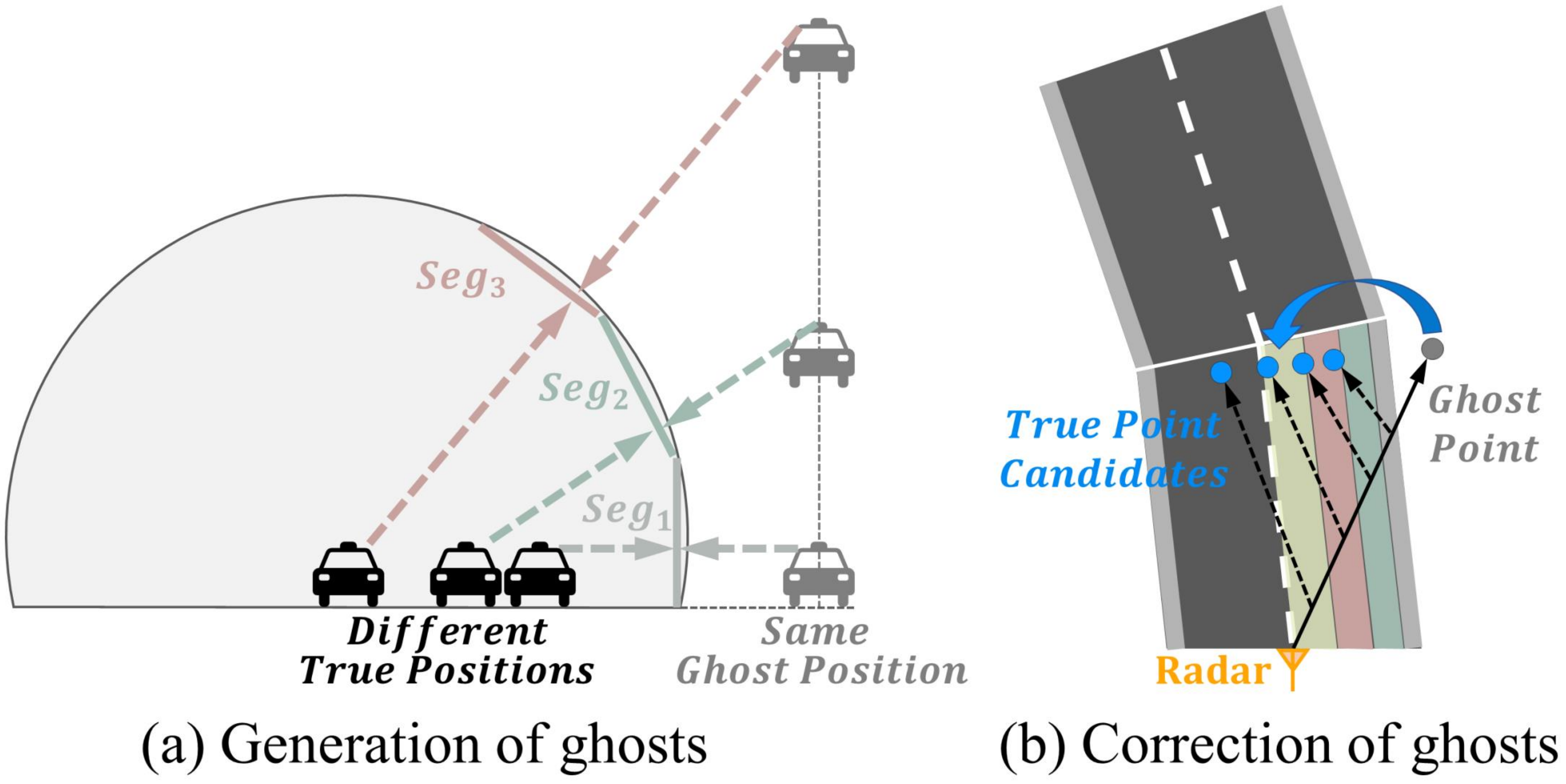}
    \caption{(a) Different true positions can lead to ghost points in the same location. (b) A ghost point may correspond to multiple true point candidates.} 
     \label{fig:gen-cor}
\vspace{-5mm}
\end{figure}

After obtaining the tunnel model, we next correct the ghost points to the true positions. To achieve this goal, we need to perform ray tracing and recover the multi-path reflections of the radar signals. However, the signals undergo complex 3D reflections within tunnel, while the radar can only capture 2D points, lacking height information. This limitation prevents us from directly recovering the 3D reflection paths. As illustrated in Figure~\ref{fig:gen-cor}(a), we observe that different true positions of a real vehicle can generate ghost points at the same location. These ghost points result from reflections off different plane segments and have varying heights. However, due to the radar's inability to capture height information, it cannot distinguish between these ghosts and, consequently, cannot directly determine the reflection paths.

Therefore, we propose a two-stage ghost correction approach. First, to bridge the gap between the 2D data and 3D reflections, we leverage the prior knowledge that real vehicles must drive on the ground plane. Using this weak constraint, we perform geometric ray tracing for the ghost points to infer the positions of the real targets. By traversing all possible reflection plane segments, we reconstruct multiple potential reflection paths for each ghost point, yielding multiple \textbf{true point candidates} (as shown in Figure~\ref{fig:gen-cor}(b)). In the second stage, we determine the final true position based on two criteria: minimizing the signal path loss along the reflection paths, and minimizing the cross-frame association distance between the candidates and the detected targets from the previous frame. The final true position is derived by merging the optimal results from both criteria.

\vspace{-3pt}
\subsubsection{True Point Candidate Generation.}
\begin{figure}
    \centering
    \setlength{\abovecaptionskip}{0.5mm}
    \includegraphics[width=0.95\linewidth]{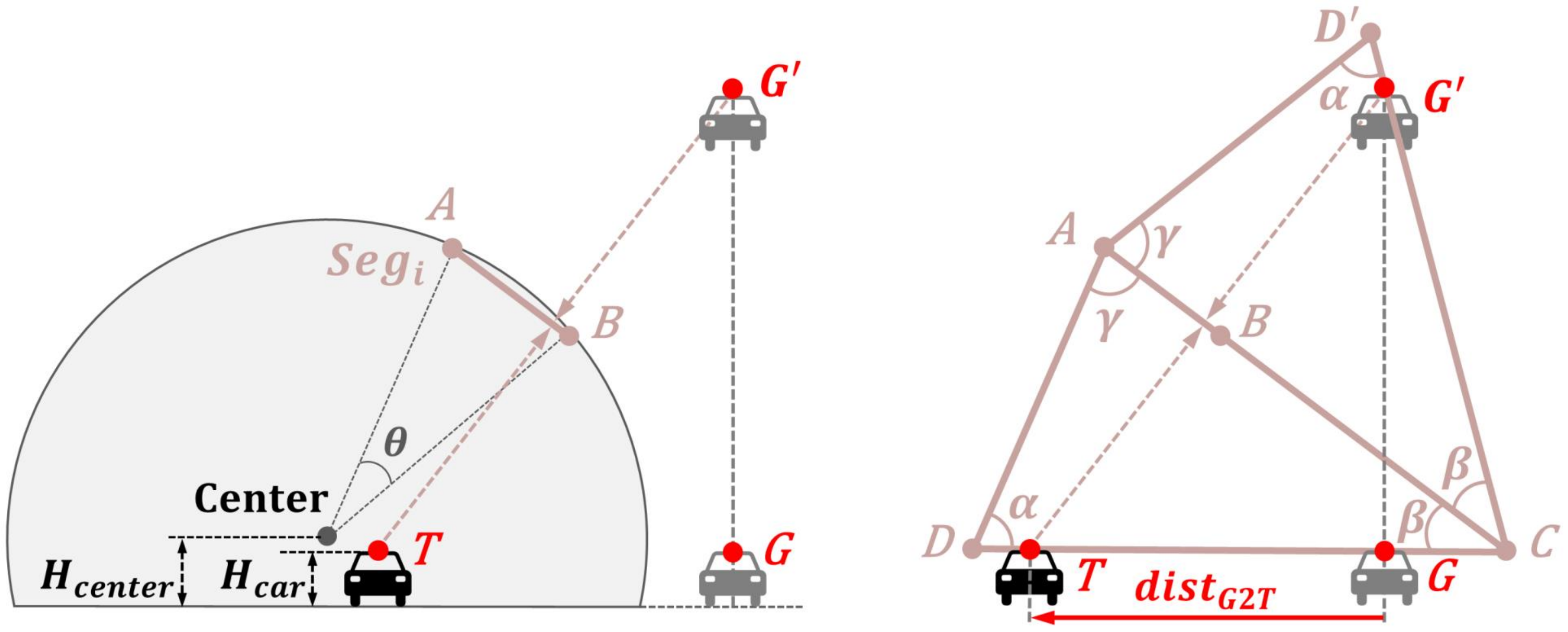}
    \caption{Illustration of the geometric relationship between the true point $T$, the ghost point $G'$, and the detected ghost point $G$. } 
     \label{fig:correction1}
\vspace{-5mm}
\end{figure}

When the radar detects a point outside the road, we connect the radar position to this point with a straight line. The plane segments that intersect this line in the radar's coordinate system are identified as potential reflection surfaces. We then traverse these possible reflection plane segments and calculate the true point candidates iteratively. To clarify this process, Figure~\ref{fig:correction1} provides an example. Notably, while real cars are driving on the ground plane, the reflection positions typically occur at the tops of cars, with their height denoted as $H_{car}$. Consequently, the true position of the car ($T$) is symmetric to the ghost position ($G'$) relative to the reflection plane ($AB$). Since the radar lacks height information, the detected ghost point is represented as $G=\left( x_g, y_g \right)$. Our goal is to determine the coordinates of true position $T=\left( x_t, y_t \right)$. By constructing two symmetrical triangles ($\triangle ACD$ and $\triangle ACD'$), we establish the geometric relationship between $T$, $G$, and $G'$. Here, point $C$ denotes the intersection of extended line $AB$ and the horizontal plane at the vehicle's roof height $H_{car}$. Point $D$ denotes the intersection of the extended line from $A$ to the tunnel's center and the same horizontal plane. Consequently, the true ghost point $G'$ must lie within the line $CD'$ under the assumption of planar reflections.

Assuming the reflection plane corresponds to the $i$-th segment in the cross-section segmentation, the angles of the triangles can be derived as follows:
\vspace{-3pt}
\begin{equation}
    \small \alpha = i\theta - \arcsin \left( \frac{H_{center}}{R_{tunnel}} \right), \; \gamma = \frac{\pi - \theta}{2}, \; \beta = \pi - \alpha - \gamma,
\end{equation}
where $\theta$ is calculated by Equation~\ref{eq:theta}.

Next, we calculate the length of side $AD$:
\vspace{-3pt}
\begin{equation}
    \small \left| AD \right| = R_{tunnel} + \frac{H_{center} - H_{car}}{\sin \alpha}.
\end{equation}

According to the sine theorem, we have:
\vspace{-3pt}
\begin{equation}
    \small \left| CD \right| = \left| AD \right| \times \frac{\sin \gamma}{\sin \beta}.
\end{equation}

We then calculate the length of $DG$:
\begin{equation}
    \small \left| DG \right| = dist_{Center} + \frac{H_{center} - H_{car}}{\tan \alpha},
\end{equation}
where $dist_{Center}$ represents the distance from the ghost point to the centerline of the tunnel. Since the tunnel path is divided into multiple straight segments, we denote the slope and intercept of the segment to which the current ghost point belongs as $m$ and $b$ (as illustrated in Figure~\ref{fig:correction2}). Then $dist_{Center}$ can be calculated as follows:
\begin{equation}\label{eq-Center}
    \small dist_{Center} = \frac{\left| mx_g + b - y_g \right|}{\sqrt{m^2 + 1}}.
\end{equation}

Finally, the distance from the ghost point to the true point ($dist_{G2T}$) can be calculated:
\vspace{-2pt}
\begin{align}\label{dist_G2T}
    \small dist_{G2T} &= \left| CT \right| - \left| CG \right| \notag \\
    &= \left| CG' \right| - \left| CG \right| \notag \\
    &= \left( \left| CD \right| - \left| DG \right| \right) \times \frac{1 - cos\left( 2\beta \right)}{cos\left( 2\beta \right)}.
\vspace{-2pt}
\end{align}
Therefore, the true point is the result of the ghost point moving $dist_{G2T}$ in the direction perpendicular to the centerline:
\vspace{-4pt}
\begin{equation}
    \small \varphi = arctan\left( \frac{1}{m} \right),
\end{equation}
\begin{equation}
    \left\{
    \begin{aligned}
        \small x_t = x_g - dist_{G2T} \times cos\varphi \\
        \small y_t = y_g + dist_{G2T} \times sin\varphi
    \end{aligned}
    \right., \; \text{if ghost point is on the right.}
\end{equation}
\begin{equation}
    \left\{
    \begin{aligned}
        \small x_t = x_g + dist_{G2T} \times cos\varphi \\
        \small y_t = y_g - dist_{G2T} \times sin\varphi
    \end{aligned}
    \right., \; \text{if ghost point is on the left.}
\end{equation}
where $\left( x_t, y_t\right)$ are the coordinates of the true point $T$.

The derivation above relies on the tunnel modeling parameters and the location of the ghost point, both of which are known. However, the parameter $H_{car}$ cannot be directly obtained by the radar. Therefore, we set $H_{car} = 1.5\,\text{m}$ during implementation, which corresponds to the average height of sedans and SUVs. Although box trucks are taller than the assumed 1.5~m height, only a small subset of high-angle reflections produce ghost points with large deviations. Our experiments show that over 85\% of ghost points from trucks are correctly corrected back within the vehicle body, supporting the practicality of this fixed-height assumption.

\subsubsection{True Position Estimation with Nearest Neighbors in Dual Domains.}

Once all true point candidates are generated, denoted as $\left( T_1,T_2, \dots , T_n \right)$, we estimate the final true position by selecting the nearest neighbors in both the signal and spatial domains. The selection follows two criteria: (1) least signal path loss first and (2) least spatial distance first. Ideally, both criteria should identify the same correct candidate, but practical challenges such as environmental noise and measurement noise introduce errors. To mitigate these issues, we integrate both selection results to refine the final true position, thereby reducing localization errors.

\begin{figure}
    \centering
    \setlength{\abovecaptionskip}{0.5mm}
    \includegraphics[width=0.82\linewidth]{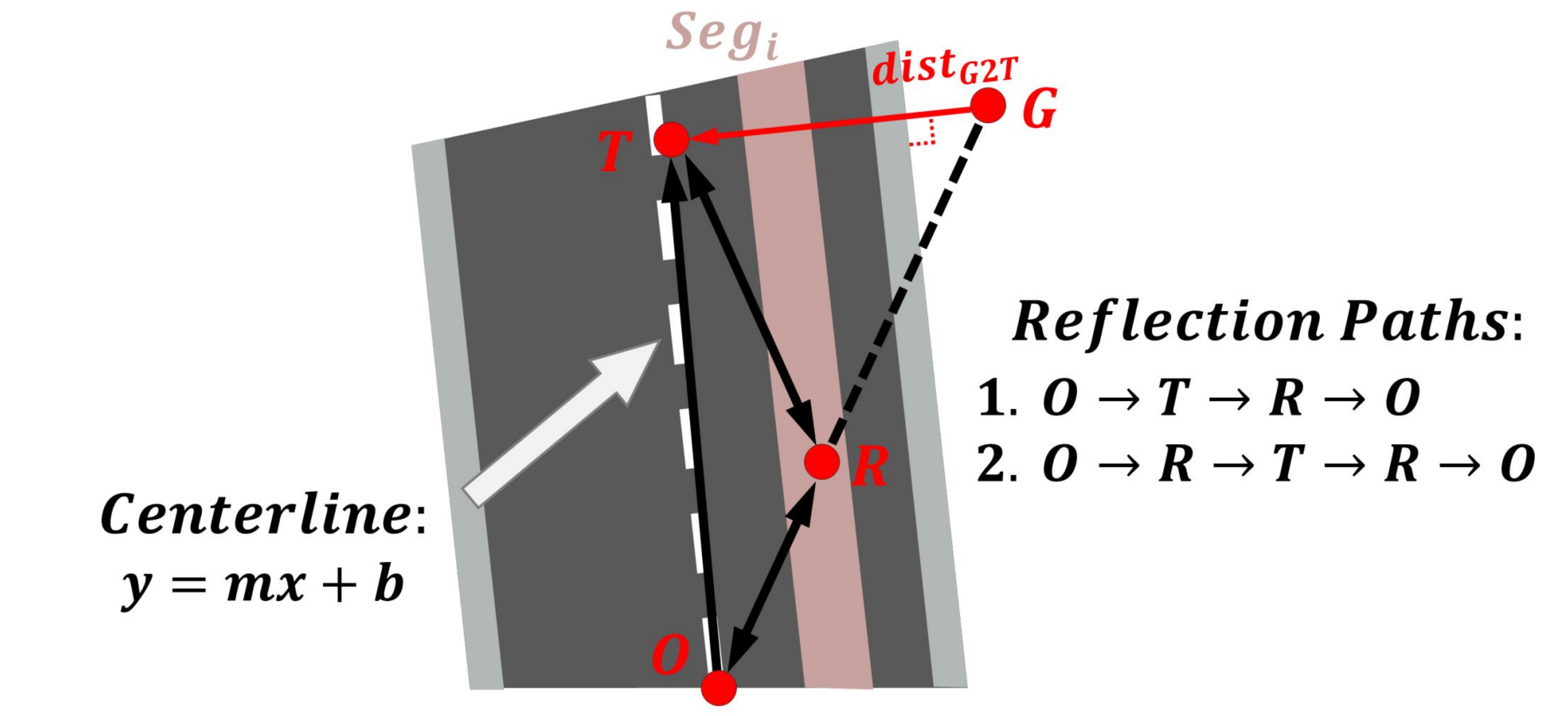}
    \caption{Illustration of the radar signals' reflection paths between the radar $O$, the true vehicle $T$, and the reflection point $R$.
    } 
     \label{fig:correction2}
\vspace{-4mm}
\end{figure}

$\bullet$ \emph{Least Signal Path Loss First.} The first criterion leverages the principle that radar points correspond to locations with maximal received power in the local region of the radar heatmap. Therefore, the reflection path which leads to the least signal path loss (i.e., the highest received power) is most likely the correct one. Figure~\ref{fig:correction2} illustrates an example of spatial signal reflections for a true point candidate (represented as $T$). The radar signals reflect between the radar $O$, the true vehicle $T$, and the reflection point $R$. Specifically, there are two probable reflection paths: $O \rightarrow T \rightarrow R \rightarrow O$ and $O \rightarrow R \rightarrow T \rightarrow R \rightarrow O$. The propagation distances along these paths are denoted as $\left| OT \right| = L_0$, $\left| OR \right| = L_1$, $\left| RT \right| = L_2$. The received radar signal power $P_r$ can be calculated according to the radar equation~\cite{radar-handbook}:
\begin{equation}
    \small P_r = \left\{
    \begin{aligned}
        \frac{P_t G_t G_r \sigma_1 \sigma_2 \lambda^2}{(4\pi)^4 (L_0 L_1 L_2)^2} \propto \frac{1}{(L_1 L_2)^2}, & \quad O \overset{L_0}{\rightarrow} T \overset{L_2}{\rightarrow} R \overset{L_1}{\rightarrow} O \\
        \frac{P_t G_t G_r \sigma_1 \sigma_2^2 \lambda^2}{(4\pi)^5 (L_1 L_2)^4} \propto \frac{1}{(L_1 L_2)^4}, & \quad O \overset{L_1}{\rightarrow} R \overset{L_2}{\rightarrow} T \overset{L_2}{\rightarrow} R \overset{L_1}{\rightarrow} O
    \end{aligned}
    \right.
\end{equation}
where $P_t$ is the transmitted power, $G_t$ and $G_r$ are the gains of the transmitting and receiving antennas, $\sigma_1$ and $\sigma_2$ are the radar cross-section (RCS) of the true vehicle and the reflection point, and $\lambda$ is the wavelength of the transmitted signal. Here, the constants $P_t$, $G_t$ and $G_r$ are determined by the radar settings. Assuming the reflectivity of radar signals is uniform across the tunnel surface, $\sigma_1$ and $\sigma_2$ can also be treated as constants, making the received power dependent solely on $L_1$ and $L_2$. Thus, the candidate that maximizes received power (i.e., minimizes path loss) is selected as:
\vspace{-6pt}
\begin{equation}
    \small T_{signal} = T_{i^*}, \enspace \text{where} \enspace i^* = \arg\max_{i} \frac{1}{L_1^i L_2^i},
\vspace{-4pt}
\end{equation}
where $L_1^i$ and $L_2^i$ represent the signal propagation distances corresponding to the $i$-th true point candidate $T_i$. 

$\bullet$ \emph{Least Spatial Distance First.} Next, we utilize the spatial information. Our idea is that the positions of true vehicles do not change abruptly within a short period (e.g., $0.1\, \text{s}$ between two consecutive frames). Therefore, the true point candidate which is closest to the detection result of the previous frame is most likely the correct one. We denote the positions of detected vehicles in the previous frame as $\mathcal{V}=\{V_1,\, V_2,\, \dots,\, V_m\}$. We then identify the true point candidate which minimizes the distance to a detected vehicle:
\vspace{-6pt}
\begin{equation}
    \small T_{dist} = T_{i^*}, \enspace \text{where} \enspace (i^*,\, j^*) = \arg\min_{i,\, j} dist(T_i,\, V_j),
\vspace{-4pt}
\end{equation}
where $dist(T_i,\, V_j)$ represents the Euclidean distance between $T_i$ and $V_j$. However, in cases where no corresponding vehicle was detected in the previous frame, all distances may be excessively large. To handle such scenarios, we introduce a distance threshold $d_{max}$, selecting $T_{dist}$ only if the minimum distance falls below $d_{max}$ (set to 4 m, corresponding to the road width in the tunnel).

\begin{figure}
    \centering
    \setlength{\abovecaptionskip}{0.5mm}
    \includegraphics[width=0.85\linewidth]{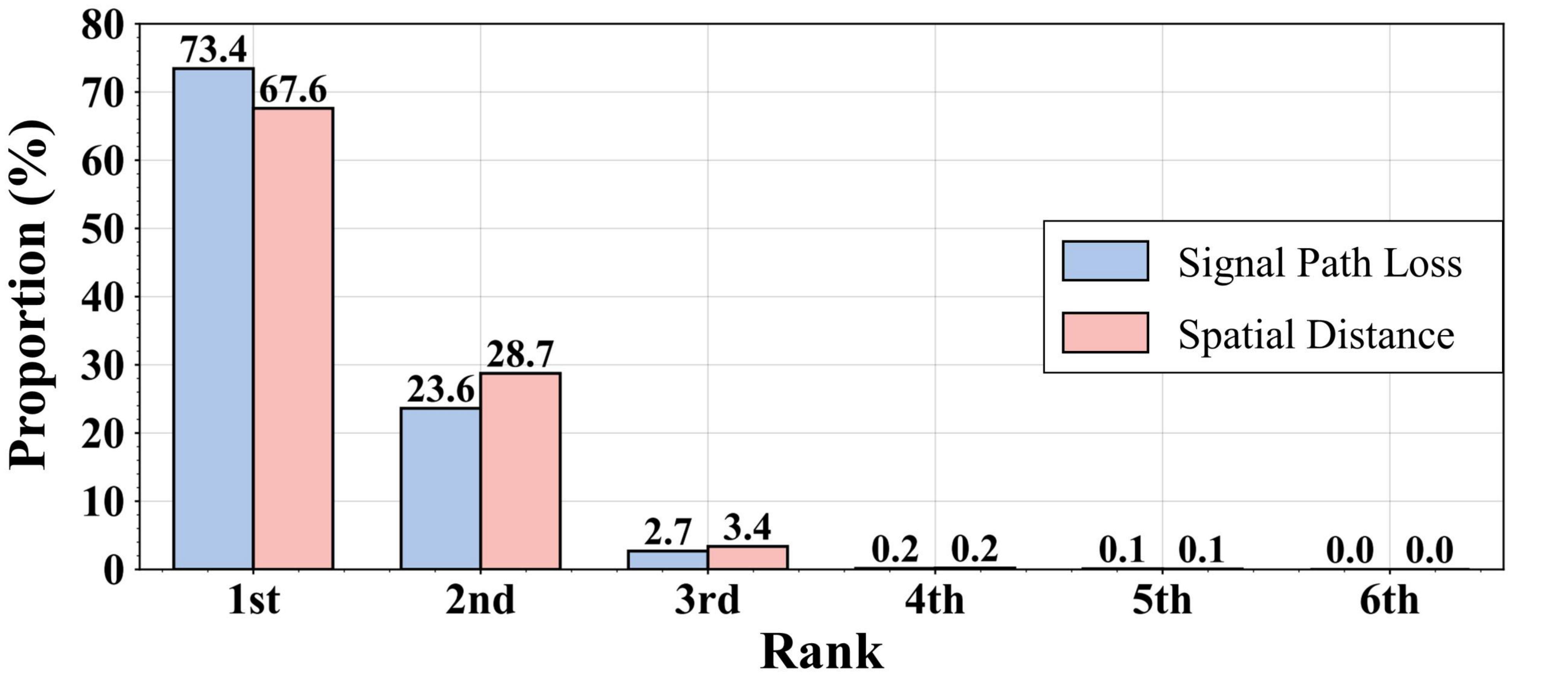}
    \caption{Proportions of correct candidates at different ranks. \textmd{The blue bars and pink bars represent rankings based on path loss and spatial distance in ascending order.}
    } 
     \label{fig:proportion}
\vspace{-4mm}
\end{figure}
$\bullet$ \emph{Final True Position Estimation.} To validate our approach, we conducted experiments in real tunnel environments and analyzed statistical results. For each ghost point's true point candidates, we first identified the correct candidate using ground truth. Then, we ranked candidates by signal path loss (blue bars in Figure~\ref{fig:proportion}) and by spatial distance (pink bars) in ascending order. As expected, the top-ranked candidates had the highest correctness proportions, aligning with our theoretical analysis. However, selection errors still occur due to factors such as uneven tunnel surface reflectivity and localization errors from previous-frame detections. To mitigate these issues, we integrate both selection results. Given candidate positions $T_{signal} = (x_1, y_1)$ and $T_{dist} = (x_2, y_2)$, we estimate the true position $(x_0, y_0)$ by minimizing the sum of squared differences between the candidate coordinates and the true position. Specifically, the objective function is formulated as:
\vspace{-3pt}
\begin{equation}
    \small \mathcal{L}(x_0, y_0) = (x_1 - x_0)^2 + (x_2 - x_0)^2 + (y_1 - y_0)^2 + (y_2 - y_0)^2 .
\vspace{-2pt}
\end{equation}
It is straightforward to demonstrate that the minimum is attained at $x_0 = \frac{x_1 + x_2}{2}$, $y_0 = \frac{y_1 + y_2}{2}$. Therefore, we define the final true position as $T_{final} = \left( T_{signal}+T_{dist} \right) / 2$. If $T_{dist}$ is not selected, we set $T_{final} = T_{signal}$. Our experimental results also demonstrate that the combined position outperforms both individual criterion, detailed in Section~\ref{sec:ablaion}.

\begin{figure*}
    \centering
    \setlength{\abovecaptionskip}{0mm}
    \includegraphics[width=0.99\linewidth]{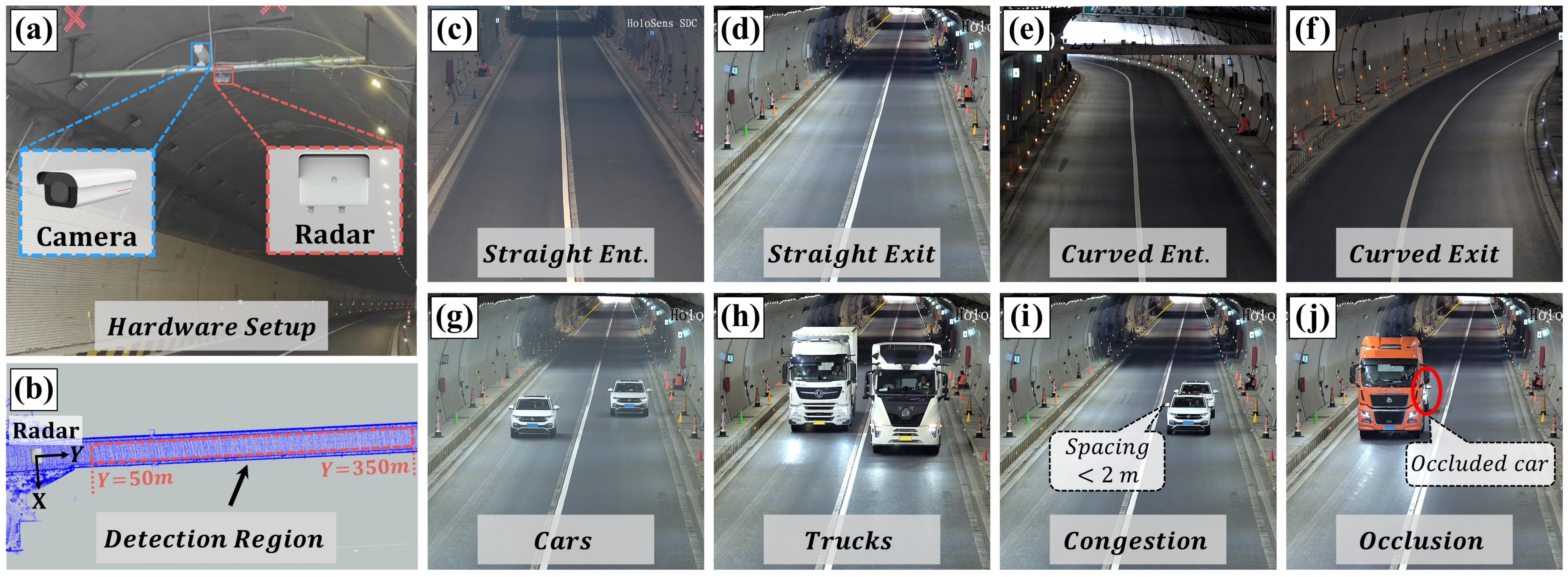}
    \caption{(a)-(b) Hardware setup and detection region of mmTunnel system. (c)-(f) Views of the entrances (Ent.) and exits of two test tunnels: Straight Tunnel and Curved Tunnel. (g)-(j) Real-world photos of the four test scenarios.} 
     \label{fig:exp-photos}
\vspace{-4mm}
\end{figure*}

\vspace{-6pt}
\subsection{Vehicle Detection}\label{vehicle-detection}
After determining the true positions of all ghost points, we next perform point clustering and target tracking to derive the accurate locations for the vehicles.

\vspace{-1mm}
\subsubsection{Point Clustering.}
We first employ the DBSCAN algorithm~\cite{dbscan}, commonly used for radar point clustering under the assumption that points belonging to a real target are spatially close to each other. DBSCAN calculates the distance between each pair of points, grouping those within a preset threshold $d_{clu}$ into the same cluster. However, due to the limited resolution of radars, points from two closely driving vehicles may be mistakenly clustered as a single target. Therefore, we incorporate Doppler velocity into the distance calculation:
\vspace{-6pt}
\begin{equation}
    \small d\left( p_i,\,p_j \right) = \sqrt{w_1\left( x_i - x_j \right)^2 + w_2\left( y_i - y_j \right)^2 + w_3\left( {v_d}_i - {v_d}_j \right)^2}
\vspace{-3pt}
\end{equation}
Based on empirical experience, we set the weights ($w_1$, $w_2$, $w_3$) = $( 1,\, 0.5,\, 4 )$, and the distance threshold $d_{clu} = 4$.

\vspace{-1mm}
\subsubsection{Target Tracking.}
We treat the cluster centers as the initial positions of vehicles and apply target tracking to further reduce localization errors. Specifically, we use a Kalman filter~\cite{kalman}, inspired by the widely adopted AB3D target-level tracking framework~\cite{ab3d}. Our tracking algorithm works as follows. For each vehicle, the Kalman filter predicts its position in the current frame based on its previous position and velocity. We assume a constant velocity motion model, which is reasonable given the high frame rate of the radar that makes velocity changes negligible between frames. Next, we compute the Euclidean distance cost matrix between the detected vehicle positions (i.e., cluster centers) and the predicted positions. Based on this matrix, we apply the Global Nearest Neighbor (GNN) algorithm to assign the matching relations between detected and predicted targets. For matched targets, the predicted positions are updated using the Kalman filter’s update step. To reduce false positives and tolerate occasional missed detections, we output a vehicle trajectory only if a target is successfully associated across three consecutive frames. Conversely, a trajectory candidate is removed if it fails to match for five consecutive frames.


\vspace{-1mm}
\section{EVALUATION THROUGH CONTROLLED EXPERIMENTS}\label{sec:evaluation}
In this section, we evaluate mmTunnel through controlled experiments. To ensure safety, these experiments were conducted in two tunnels that are not open to the public. Furthermore, we conducted experiments in a public tunnel with real-world traffic, which will be presented in Section~\ref{sec:real-world}.

\vspace{-3mm}
\subsection{Experimental Setup}
\textbf{System Implementation.} We implemented our system in real tunnels using a traffic radar (ASN850~\cite{asn850}) and a camera (M2391-10-TL~\cite{M2391}). The devices were installed at the entrances or exits of the tunnels at a height of $5.1\,\text{m}$. Note that the camera is not part of our system and is used solely for verification purposes. For controlled testing, we selected two tunnels that are not open to the public: a straight tunnel (hereafter referred to as the \emph{Straight Tunnel}) and a curved tunnel (hereafter referred to as the \emph{Curved Tunnel}). To evaluate the system’s performance under different tunnel environments and vehicle driving directions, we installed the devices sequentially at the entrances and exits of both tunnels, conducting experiments at each location. The radar’s sensing range covered 50~m to 350~m from its deployment location, with the first 50~m reserved as a blind zone. The hardware setup and tunnel environments are shown in Figure~\ref{fig:exp-photos}(a)-(f).

\vspace{1pt}\noindent\textbf{Radar Calibration and Tunnel Modeling.} After installing the devices, we calibrated the radar using three corner reflectors. Specifically, the reflectors were placed at equal intervals across the road width (left, center, and right) and moved forward in steps of $10\,\text{m}$, starting from $50\,\text{m}$ to $350\,\text{m}$ from the radar location. This process resulted in 31 sets of measurements, providing 93 sampled positions. Using the radar points corresponding to the reflectors, we calibrated the radar and fitted the tunnel centerline equation. Additionally, the two tunnels' cross-sections are nearly identical (see Figure~\ref{fig:segment1}), with parameters of $R_{tunnel}=5.5\,\text{m}$, $W_{road}=4\,\text{m}$, and $H_{center}=1.6\,\text{m}$. Tunnel modeling was then carried out using the method described in Section~\ref{tunnel-modeling}.

\vspace{1pt}\noindent\textbf{Test Scenarios.} To evaluate our system under diverse real-world conditions, we designed four test scenarios, as shown in Figure~\ref{fig:exp-photos}(g)-(j):
\vspace{-3pt}
\begin{itemize}[label={$\bullet$}, labelsep=0.5em, left=0em, itemindent=0em]
    \item \emph{Cars:} One or two cars driving in the tunnel.
    \item \emph{Trucks:} One to three trucks driving in the tunnel.
    \item \emph{Congestion:} Two cars driving slowly with their inter spacing below $2$ meters, simulating congestion in the tunnel.
    \item \emph{Occlusion:} A truck and a car driving in the same lane, with the car almost fully occluded by the truck.
    \vspace{-2pt}
\end{itemize}

\noindent We deployed our system sequentially at the entrances and exits of both tunnels, and conducted experiments of the four scenarios at each location. For each scenario, we performed six data collection sessions, each capturing the entire process from the vehicles entering the tunnel to their exit. In total, we obtained 46,720 frames of valid data at 10 FPS.

\begin{figure*}
    \centering
    \setlength{\abovecaptionskip}{0mm}
    \includegraphics[width=0.88\linewidth]{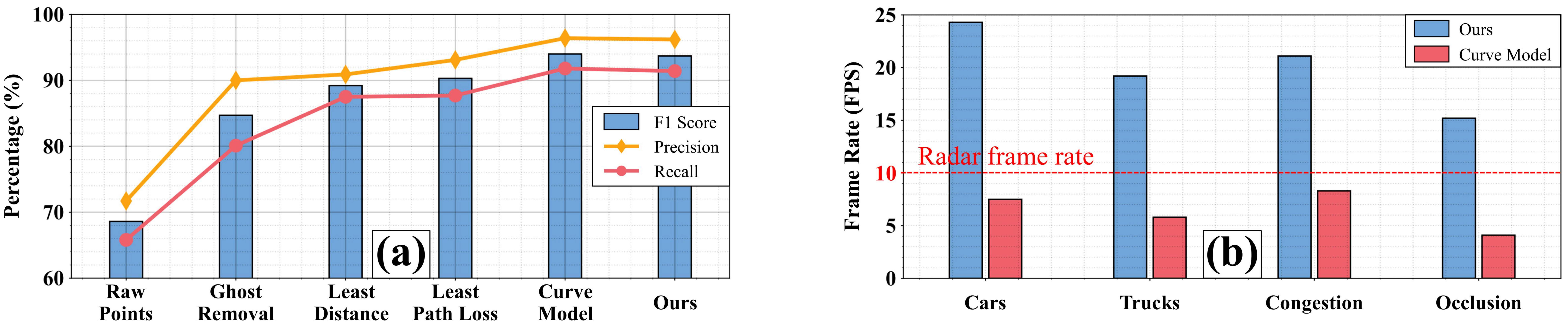}
    \caption{Overall detection performance and processing speed of mmTunnel. \textmd{(a) Comparison of mmTunnel with other methods in terms of vehicle detection performance. (b) Processing frame rate of system with our segmentation-based modeling and curved-surface modeling on an Intel Core i5-11500 processor. }}
    \label{fig:ablation-fps}
    \vspace{-3mm}
\end{figure*}

\vspace{1pt}\noindent\textbf{Ground Truth.} To obtain accurate ground-truth locations of vehicles, we deployed two 80-line LiDARs (RS-Ruby Lite~\cite{RS-Ruby}) inside the tunnel. Each LiDAR provides precise vehicle detection within a 120 m radius. To ensure full coverage, the LiDARs were installed on the tunnel walls---one on the left and one on the right---positioned 150 m and 250 m away from the radar deployment location, respectively. This setup enabled the two LiDARs to collaboratively capture the bounding boxes of all vehicles throughout the tunnel.

\vspace{1pt}\noindent\textbf{Performance Metrics.} We first perform minimum-distance binary matching between the detected targets and ground truth. If the $x$-axis and $y$-axis distances between a matched pair are both within predefined thresholds, the detection is counted as a true positive ($TP$). Unmatched detections, as well as matched pairs with distances exceeding the thresholds, are counted as false positives ($FP$). Similarly, unmatched ground-truth objects or those in matched pairs exceeding the thresholds are counted as false negatives ($FN$). The $x$-axis threshold is set to $1.5$ m to ensure lane-level localization accuracy, while the $y$-axis threshold is set to $5$ m to account for the size of large trucks, which can exceed $10$~m in length. The performance of vehicle detection is evaluated by precision $P$, recall $R$, and $F1$ score (the harmonic mean of $P$ and $R$): $P = \frac{TP}{TP + FP}$, $R = \frac{TP}{TP + FN}$, $F1 = \frac{2 \times P \times R}{P+R}$.

\vspace{-2mm}
\subsection{Vehicle Detection Performance}
\subsubsection{Overall Performance.} \label{sec:ablaion}
We first compared the overall performance of mmTunnel with five other methods, specifically designed for comparison purposes:
\begin{enumerate}[left=0pt]
\vspace{-2pt}
    \item \emph{Raw Points,} in which vehicle detection is performed directly using the raw radar points.
    \item \emph{Ghost Removal,} in which ghost points located outside the driving lanes are removed before vehicle detection.
    \item \emph{Least Distance,} in which only the least spatial distance criterion is applied to select the optimal true point.
    \item \emph{Least Path Loss,} in which only the least signal path loss criterion is applied to select the optimal true point.
    \item \emph{Curve Model,} in which the segmentation-based modeling method is replaced with curved-surface modeling.
\end{enumerate}
\vspace{-2pt}
Notably, the last three methods replace specific components in the mmTunnel system, and the remaining modules are kept the same. The results are presented in Figure~\ref{fig:ablation-fps}(a). Vehicle detection using raw radar points yields the lowest performance, with an F1 score of 68.6\%. While filtering out ghost points improves detection to some extent, it does not address the loss of radar points corresponding to true vehicle positions, particularly in occluded scenarios. This limitation leads to frequent missed detections, resulting in a low recall of 80.1\%. Our ghost point correction method significantly enhances performance, but relying on only one criterion (either spatial distance or signal path loss) yields F1 scores that are 4.6\% and 3.4\% lower, respectively, compared to the combined approach. Furthermore, the segmentation-based modeling performs comparably to curved-surface modeling, with only a 0.3\% decrease in F1 score, indicating that our segmentation process barely introduces additional errors. Overall, our system achieves both efficient and accurate vehicle detection, achieving an F1 score of 93.7\%.

\begin{figure}
    \centering
    \setlength{\abovecaptionskip}{0mm}
    \includegraphics[width=0.99\linewidth]{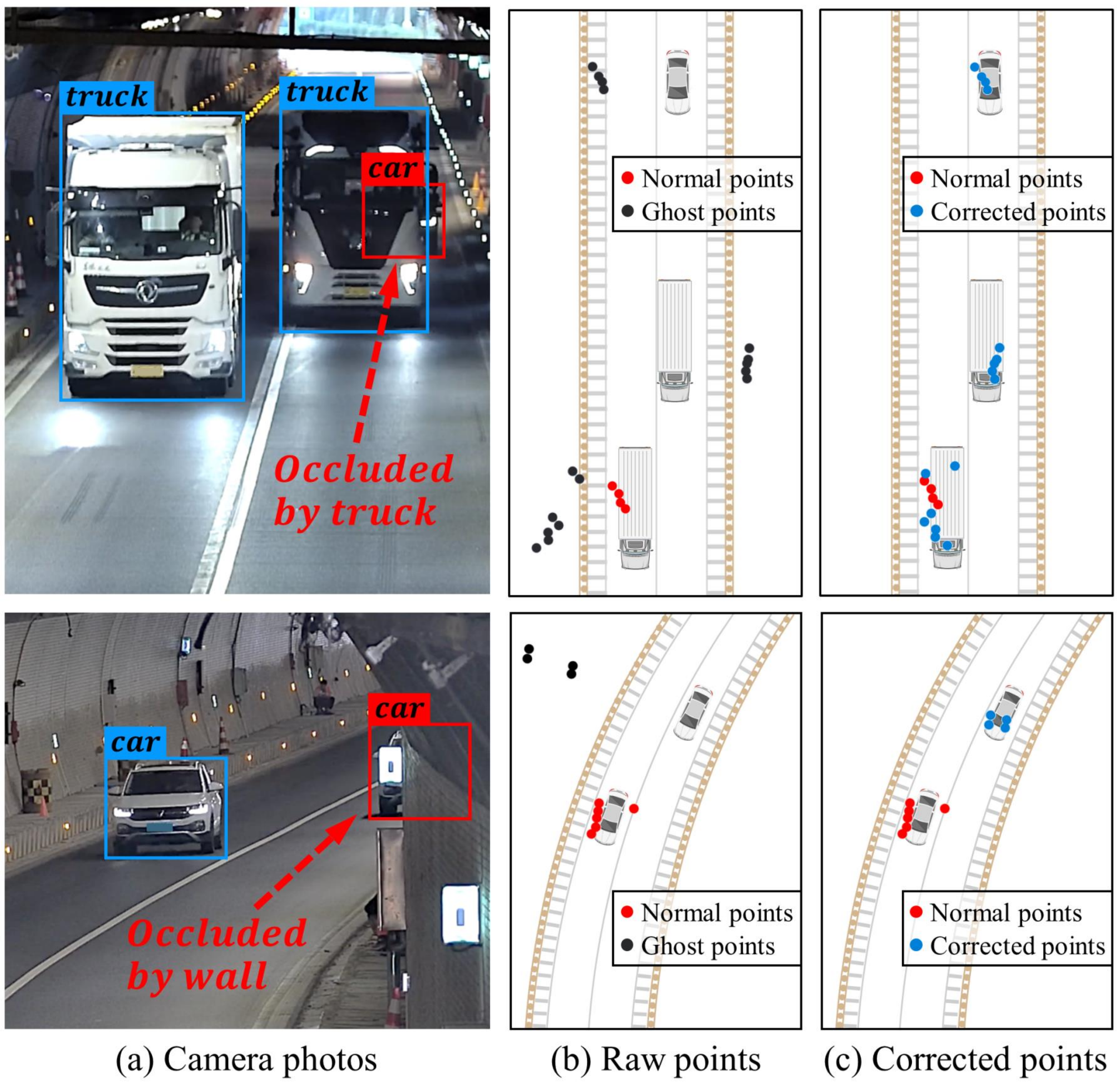}
    \caption{Visualization results from both the Straight Tunnel and Curved Tunnel. \textmd{Our system effectively corrects ghost points to true positions, enabling accurate vehicle detection even in scenarios where cars are fully occluded.}}
     \label{fig:visualize}
\vspace{-4mm}
\end{figure}

\vspace{-2mm}
\subsubsection{Processing Speed.}
In addition to detection accuracy, processing speed is a critical factor for real-time systems. We evaluated the frame rates of our system using segmentation-based modeling and curved-surface modeling across the four test scenarios, as shown in Figure~\ref{fig:ablation-fps}(b). The processing was conducted on an Intel Core i5-11500 processor. Here, the radar frame rate refers to the radar's data output rate (i.e., 10~FPS), while the system frame rate indicates the processing speed of our system when executed offline. Therefor, mmTunnel must achieve a processing frame rate exceeding 10~FPS to ensure real-time operation.

The results demonstrate that segmentation-based modeling consistently satisfies the real-time processing requirement across all scenarios. In contrast, curved-surface modeling requires solving complex equations for multi-path reflection tracing, leading to significantly lower frame rates. Notably, segmentation-based modeling achieves a $2.5 \sim 3.7$ times higher frame rate compared to curved-surface modeling, highlighting its efficiency for real-time applications.


\vspace{-1mm}
\subsubsection{Visualization Results.}
Figure~\ref{fig:visualize} illustrates the results of mmTunnel, accompanied by corresponding camera photos for reference. Panel (a) shows real-world scenarios in the camera view. Panel (b) illustrates the raw radar point cloud, which consists of both normal points (red) and ghost points (black). Panel (c) displays the corrected radar points (blue) after applying our ghost point correction method, aligning them with the true vehicle positions.

As shown, ghost points frequently occur in tunnels, resulting in significant vehicle localization errors. However, our system effectively corrects these ghost points in both Straight Tunnel and Curved Tunnel scenarios, enabling accurate vehicle localization. More importantly, even when vehicles are entirely occluded by other vehicles or tunnel walls, our system leverages the corrected ghost points to detect and localize them accurately. Quantitatively, over 80\% of ghost points are relocated within the corresponding vehicle bounding boxes after correction, further demonstrating the effectiveness of our algorithm.

\begin{figure}
    \centering
    \setlength{\abovecaptionskip}{0mm}
    \includegraphics[width=0.99\linewidth]{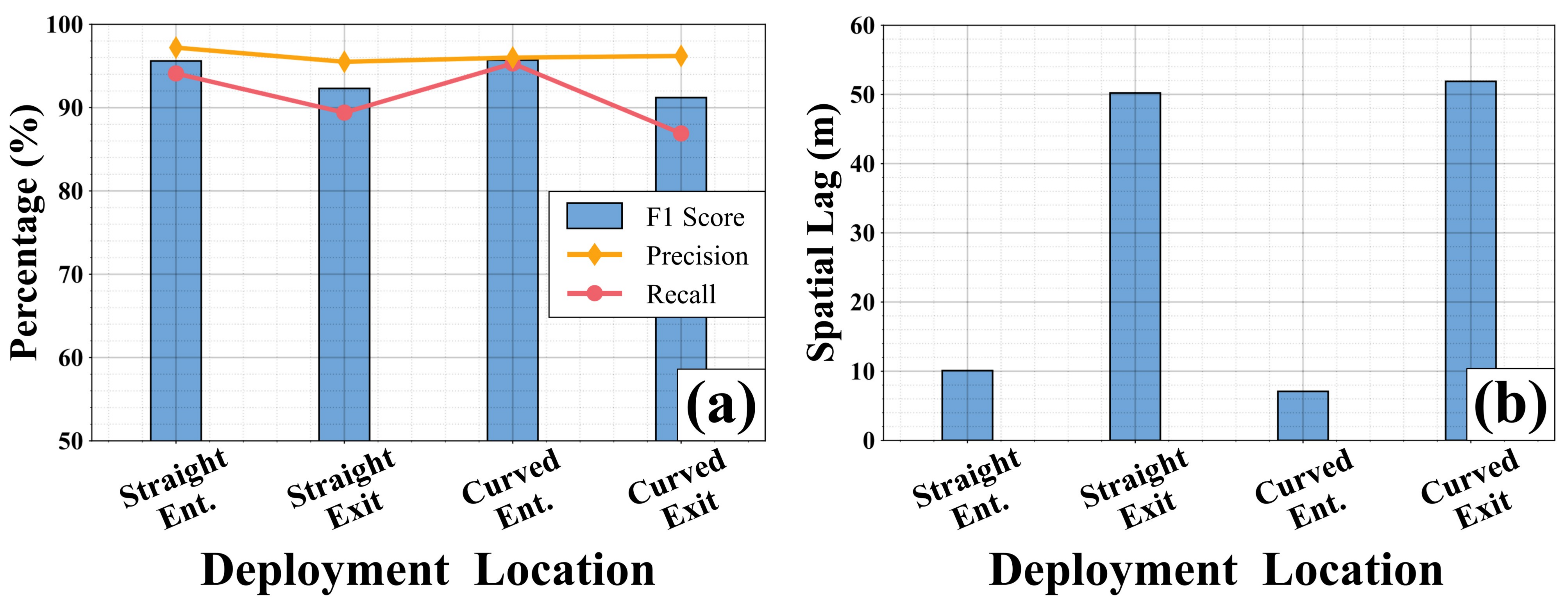}
    \caption{(a) Vehicle detection performance, and (b) average spatial lag of detected vehicle trajectories under different radar deployment locations.} 
     \label{fig:deploy}
\vspace{-4mm}
\end{figure}

\vspace{-1mm}
\subsubsection{Impact of Deployment Location.}\label{sec:deploy}
The radar deployment location is another critical factor to consider. As shown in Figure~\ref{fig:deploy}(a), the F1 score is 3.3\% to 4.5\% lower when the radar is installed at the tunnel entrance compared to the exit. This performance difference is primarily due to two factors. First, the radar's fixed angular resolution results in greater lateral localization errors for distant targets compared to those closer to the radar. Second, the Kalman filter-based tracking algorithm relies on associating raw clustering results across frames, and its ability to generate accurate trajectories depends heavily on the stability of these input clusters. Therefore, when the radar is installed at the tunnel entrance, vehicles move from near to far relative to the radar. In this configuration, closer targets are detected with higher accuracy, and the tracker can rapidly generate stable trajectories, leading to improved overall performance. Conversely, when the radar is placed at the tunnel exit, vehicles approach the radar from far to near. In this scenario, larger localization errors and less stable clustering results for distant targets hinder the tracker’s ability to produce stable trajectories.

Figure~\ref{fig:deploy}(b) presents the average \emph{spatial lag} of detected vehicle trajectories under four different radar deployment positions. Here, spatial lag refers to the distance between the location where a vehicle’s trajectory is first generated and the starting boundary of the radar’s detection region. A larger spatial lag indicates that a vehicle is detected later, leaving the earlier portion of its path undetected. The results show that when the radar is placed at the tunnel entrance, most vehicle trajectories are generated promptly, with an average spatial lag of around 10 meters. In contrast, at the tunnel exit, the spatial lag increases to about 50 meters, meaning a larger portion of vehicle trajectories experience delayed detection, leading to a reduced recall rate. These findings highlight that aligning the radar’s sensing direction with the direction of vehicle movement in the tunnel improves performance by enabling earlier detection of vehicles.

\begin{figure}
    \centering
    \setlength{\abovecaptionskip}{0mm}
    \includegraphics[width=0.99\linewidth]{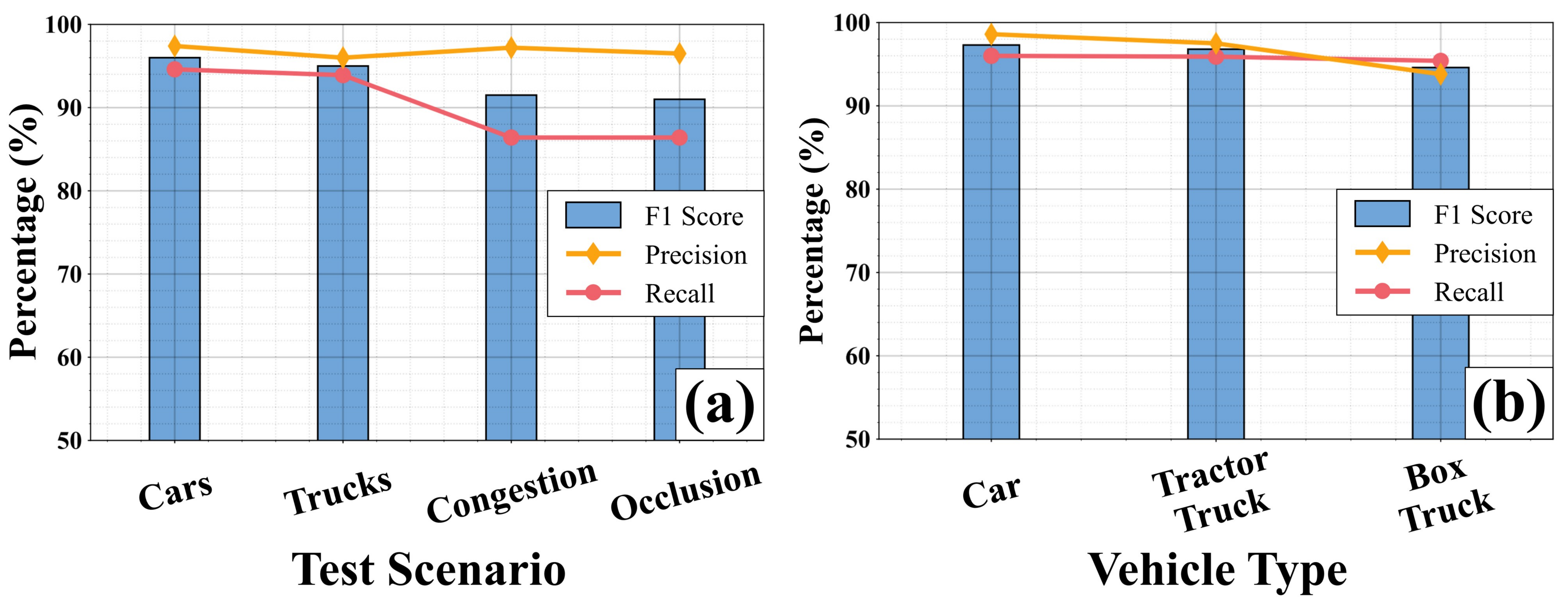}
    \caption{(a) Vehicle detection performance under the four test scenarios. (b) Vehicle detection performance across different vehicle types.} 
    \label{fig:scenario}
\vspace{-4mm}
\end{figure}

\vspace{-1mm} 
\subsubsection{Performance across Test Scenarios.}
The performance of our system is influenced by the specific test scenarios, which vary based on vehicle types and driving conditions. Figure~\ref{fig:scenario}(a) illustrates the system's performance across the four test scenarios. The results show that the recall rate for the \emph{Congestion} scenario is 8.2\% lower compared to the normal \emph{Cars} scenario. This drop is expected, as the two cars in the \emph{Congestion} setting drive closely (inter-vehicle distance $<2$~m) at speeds below 10~km/h. Thus, the radar points are near each other in both position and Doppler velocity, occasionally causing them to be mistakenly detected as a single target. Nevertheless, the \emph{Congestion} scenario still achieves an F1 score of 91.5\%, demonstrating the system’s robustness in dense traffic conditions.

Additionally, when a car is occluded by a truck, the recall rate drops by 7.5\% compared to the normal \emph{Trucks} scenario. This is expected since the occluded car can only be detected using corrected ghost points, which are not always stable, resulting in occasional missed detections. Ours experiments show that the car remains occluded (i.e., invisible to the camera) for over 73\% of the time in this scenario. Despite the challenge, the F1 score remains above 91\%, indicating the system's ability to detect vehicles under long-term occlusion.

\vspace{-1mm} 
\subsubsection{Impact of Vehicle Type.}
The type of vehicle significantly affects the distribution of radar points and, consequently, the detection of vehicles due to their varying shapes and sizes. To evaluate this, we compared detection performance across three categories: cars, tractor trucks, and box trucks. As shown in Figure~\ref{fig:scenario}(b), the F1 score for box trucks is 2.2\% to 2.7\% lower than that for the other types. This is primarily because box trucks are typically large and metallic, causing more complex radar signal reflections. These reflections can generate ghost points that are challenging to correct, resulting in localization errors. Despite this, the F1 score for detecting box trucks remains above 94\%, demonstrating the robustness of mmTunnel in detecting various vehicle types within tunnels. Note that we only use data from the \emph{Cars} and \emph{Trucks} scenarios, to evaluate the impact of vehicle type without interference from other conditions.

\begin{figure}
    \centering
    \setlength{\abovecaptionskip}{0mm}
    \includegraphics[width=0.99\linewidth]{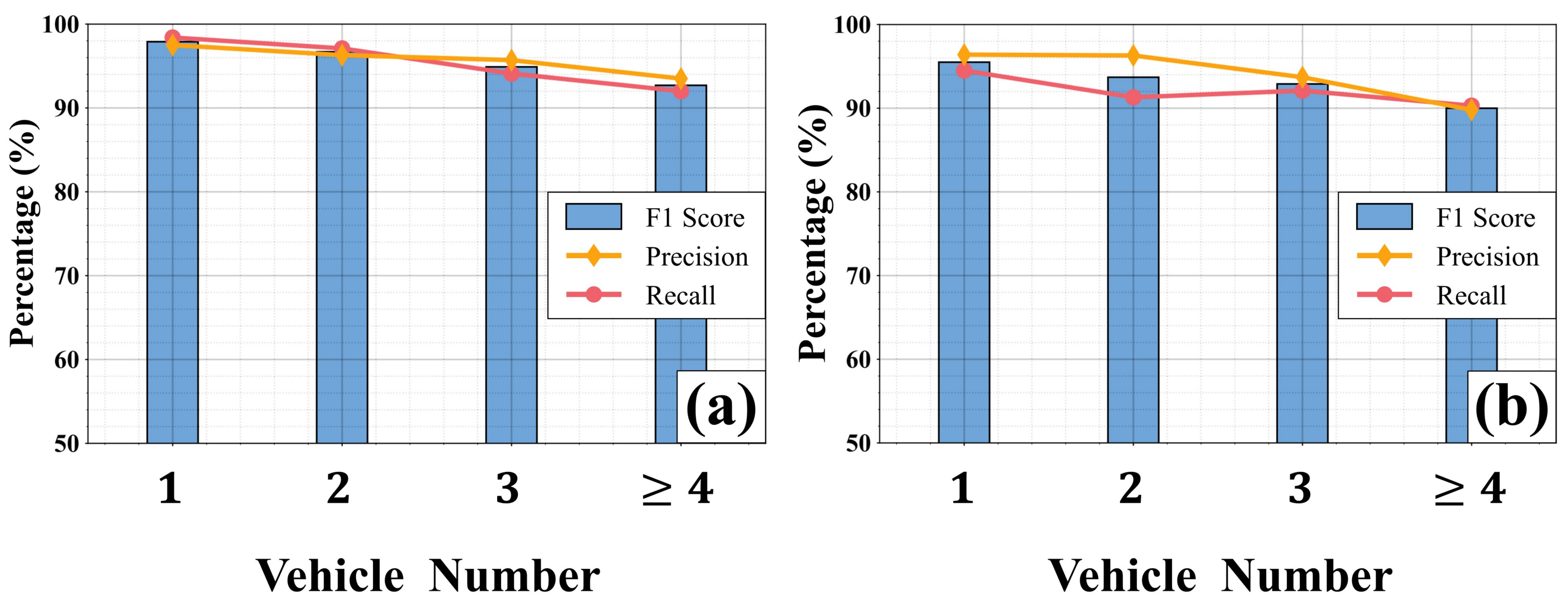}
    \caption{Vehicle detection performance across different number of vehicles with radar deployed at (a) the tunnel entrance, and (b) the tunnel exit.} 
     \label{fig:vehicle_number}
\vspace{-4mm}
\end{figure}

\vspace{-1mm} 
\subsubsection{Impact of Vehicle Number.}
The number of vehicles in the tunnel also affects the system’s performance. Figure~\ref{fig:vehicle_number} compares vehicle detection performance under different vehicle densities for radar deployed at the tunnel entrance (Figure~\ref{fig:vehicle_number}(a)) and exit (Figure~\ref{fig:vehicle_number}(b)). As the number of vehicles increases, the F1 score for vehicle detection decreases slightly. This decline is primarily due to the interference caused by dense radar points, which complicates the detection process, as well as the increased frequency of vehicle occlusion. Nevertheless, even with four or more vehicles present, the system maintains an F1 score of 90\%, demonstrating its robustness in high-traffic tunnel scenarios.

\begin{figure}
    \centering
    \setlength{\abovecaptionskip}{0mm}
    \includegraphics[width=0.99\linewidth]{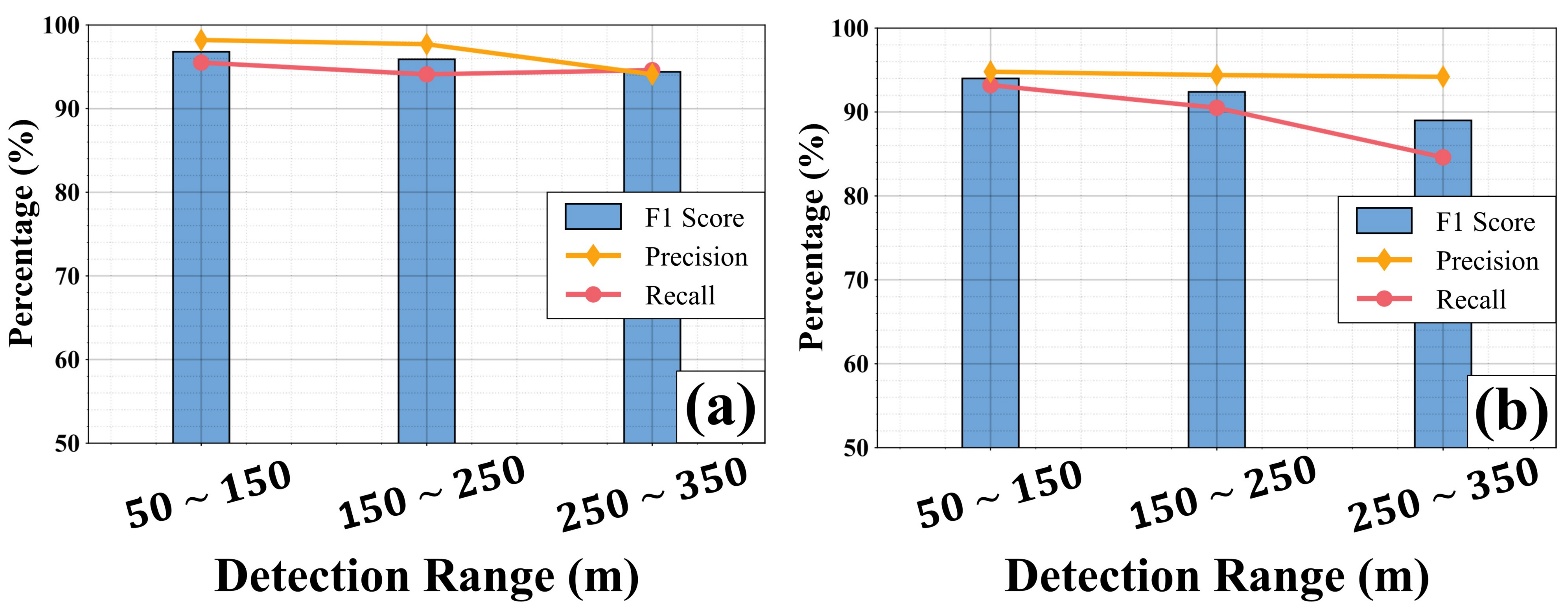}
    \caption{Vehicle detection performance across different detection ranges with radar deployed at (a) the tunnel entrance, and (b) the tunnel exit.} 
     \label{fig:range}
\vspace{-4mm}
\end{figure}

\vspace{-1mm} 
\subsubsection{Impact of Detection Range.}
The mmTunnel system is designed to maintain high performance within a sensing range of 50 m to 350 m from the radar's deployment location. Figure~\ref{fig:range} compares vehicle detection performance across different sensing ranges for radar deployed at the tunnel entrance (Figure~\ref{fig:range}(a)) and exit (Figure~\ref{fig:range}(b)). The results show that detection performance declines as the sensing range increases, with a more pronounced drop when the radar is deployed at the tunnel exit. This difference, as previously analyzed in Section~\ref{sec:deploy}, is attributed to the movement direction of vehicles relative to the radar. However, when the radar is deployed at the tunnel entrance, the F1 score consistently exceeds 94\% across all sensing ranges, demonstrating the system's robustness in far-range detection.
\vspace{-1mm}
\section{EVALUAION WITH REAL-WORLD TRAFFIC} \label{sec:real-world}
After conducting extensive controlled experiments, we next assess the performance of mmTunnel in a public tunnel with real-world traffic.

\vspace{-2mm}
\subsection{Experimental Setup}
\begin{figure}
    \centering
    \setlength{\abovecaptionskip}{0mm}
    \includegraphics[width=0.99\linewidth]{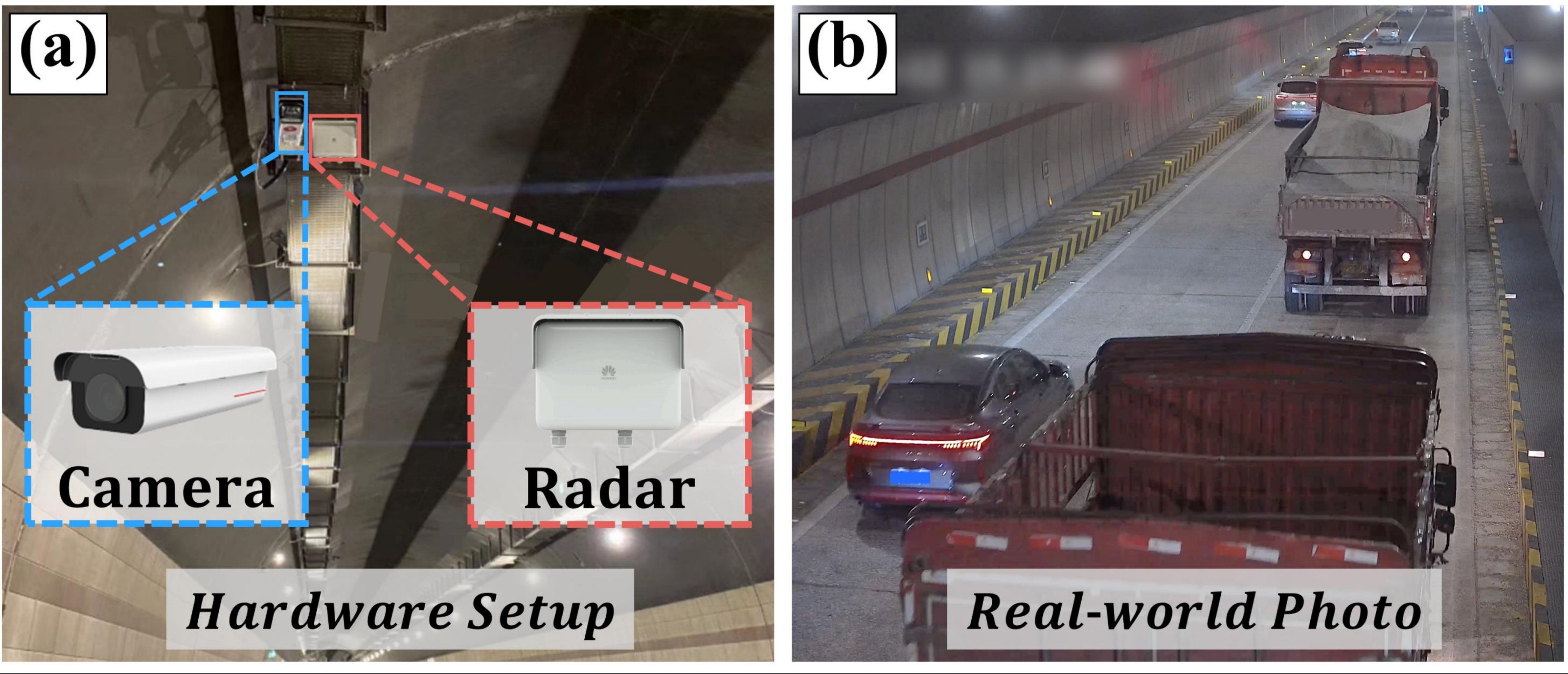}
    \caption{(a) Hardware setup of mmTunnel system in the public tunnel. (b) Real-world photo of the dense traffic in tunnel.} 
     \label{fig:changan_photo}
\vspace{-6mm}
\end{figure}

Due to differences in the tunnel's infrastructure, the devices in this experiment were installed on the upper side of the tunnel wall, rather than on the middle of the tunnel ceiling as in the controlled experiments. The hardware setup and tunnel environment are shown in Figure~\ref{fig:changan_photo}. Note that the camera is not part of our system and is used solely for verification purposes. To enhance detection performance, we aligned the radar's sensing direction with the direction of vehicle movement in the tunnel. We also conducted radar calibration and tunnel modeling. The parameters of the tunnel cross-section are as follows: $R_{tunnel}=5.6\,\text{m}$, $W_{road}=4.4\,\text{m}$, and $H_{center}=2\,\text{m}$.

\vspace{-2mm}
\subsection{Dataset}
We recorded data at different times, generating five sequences (Seq1 $\sim$ Seq5), with each sequence containing over 7,200 frames. The traffic in the tunnel was consistently dense, resulting in a final dataset containing more than 36,000 frames (over 60 minutes) and 110,000 vehicle samples.

Our camera (M2391-10-TL~\cite{M2391}) provides high-resolution videos and image-based detection results, which were used for verification purposes. These data also allowed us to identify and quantify the types of vehicles in our dataset. Specifically, our dataset includes 448 cars, 59 flatbed trucks, and 82 box trucks. The frequent presence of trucks results in more occlusions, making vehicle detection more challenging. Additionally, we counted the number of vehicles within the radar's sensing range at each frame to provide an intuitive understanding of the traffic density in the tunnel, as shown in Figure~\ref{fig:changan_vehicle_num}. More than half of the time, there were at least 3 vehicles simultaneously within the radar's sensing range, with up to 12 vehicles at most. Such high vehicle density further increases the challenge of accurate vehicle detection.

\subsection{Vehicle Detection Performance}
\subsubsection{Overall Performance.}
\begin{figure}[t]
    \centering
    \begin{minipage}[t]{0.48\linewidth}
        \centering
        \setlength{\abovecaptionskip}{0mm}
        \includegraphics[width=\linewidth]{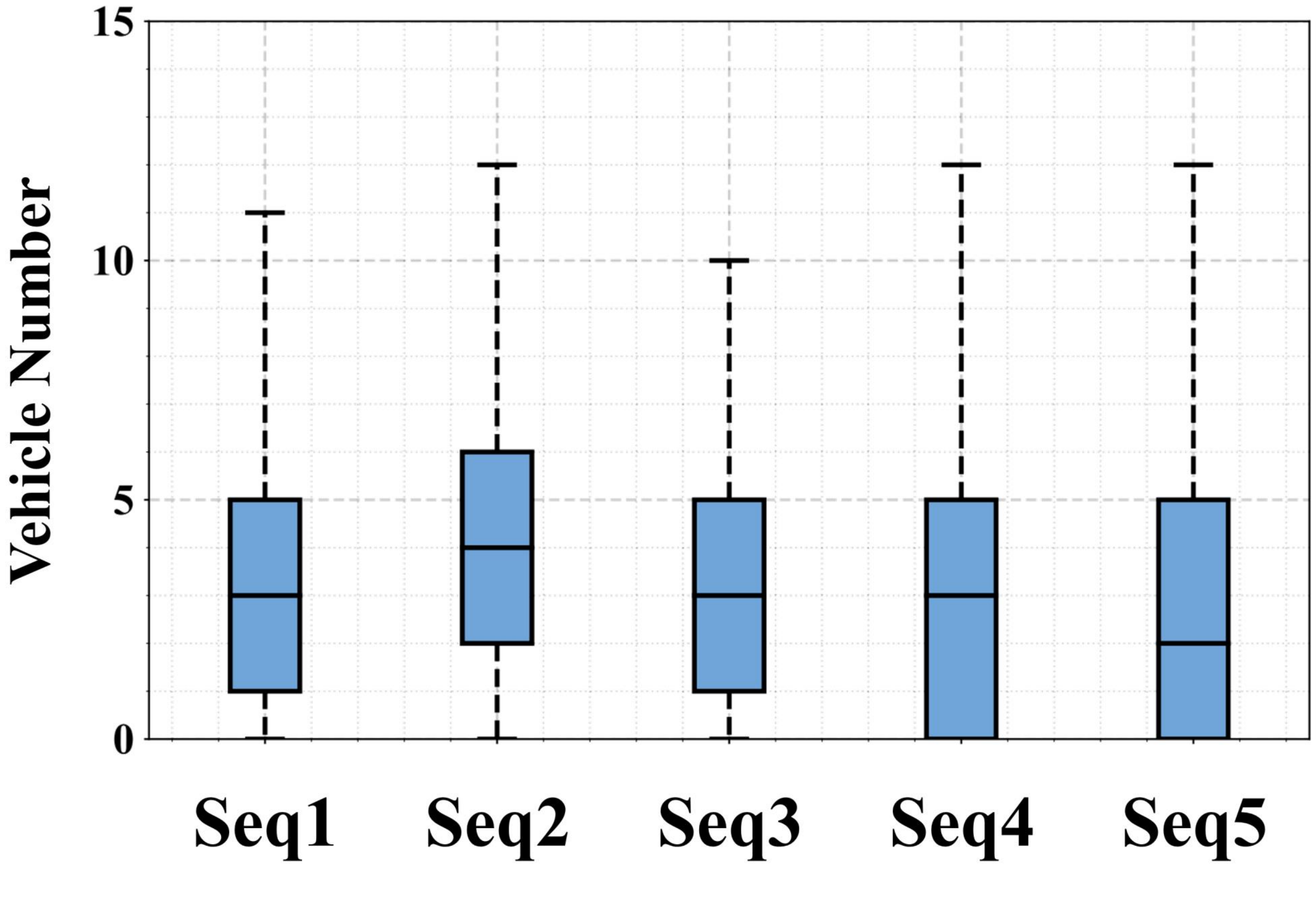}
        \caption{Number of vehicles in radar's sensing range across 5 sequences.}
        \label{fig:changan_vehicle_num}
    \end{minipage}
\hspace{1mm}
    \begin{minipage}[t]{0.48\linewidth}
        \centering
        \setlength{\abovecaptionskip}{0mm}
        \includegraphics[width=\linewidth]{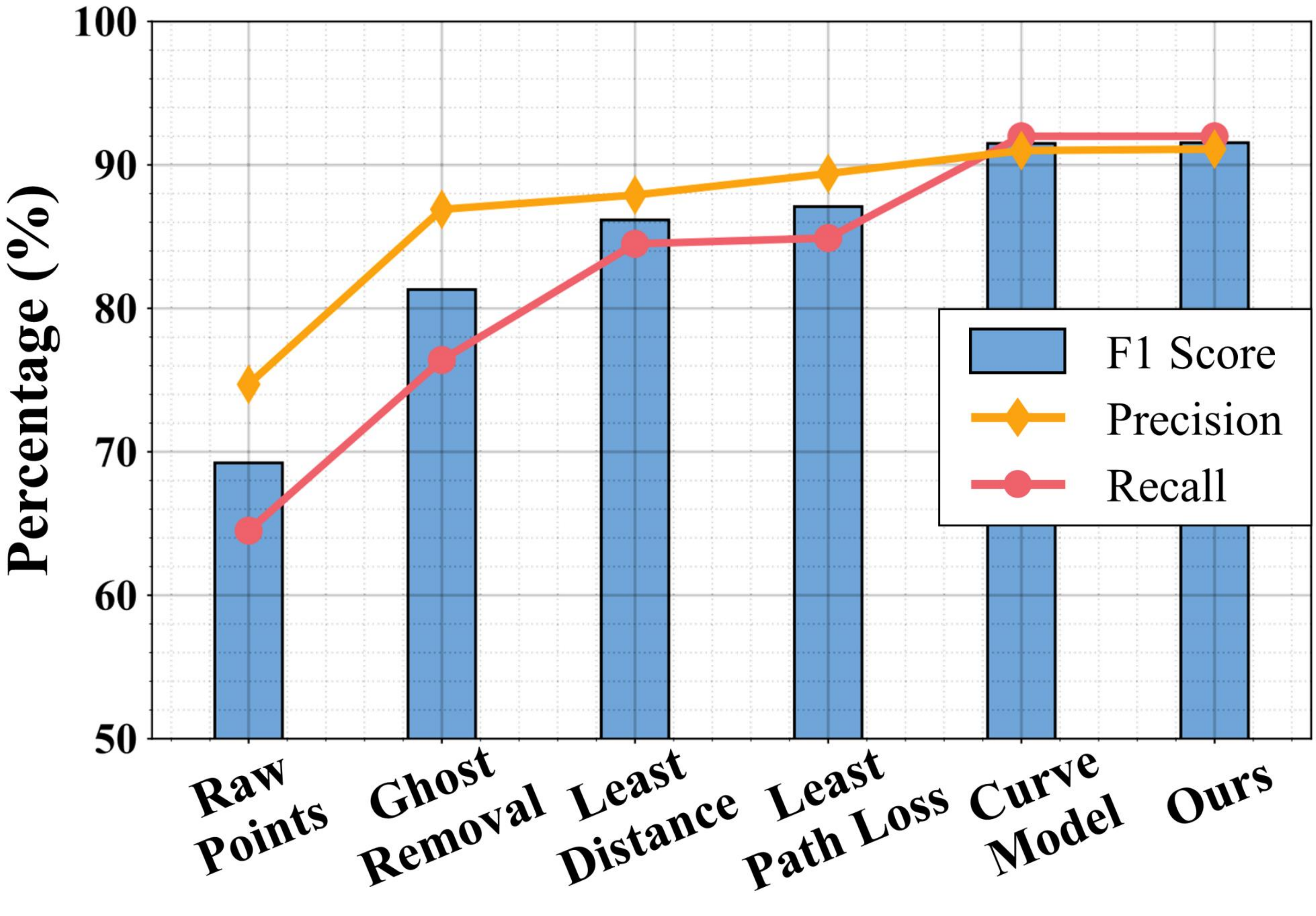}
        \caption{Comparison of mmTunnel with other methods in public tunnel.
        }
        \label{fig:changan_ablation}
    \end{minipage}
\vspace{-3mm}
\end{figure}

We first compared the overall performance of mmTunnel with other methods to assess the effectiveness of our methods. The settings are the same as Section~\ref{sec:ablaion}. The results are presented in Figure~\ref{fig:changan_ablation}. Vehicle detection using raw radar points yields the lowest performance, with an F1 score of 69.2\%. Filtering out ghost points improves detection to some extent, but the F1 score of 81.3\% is still not satisfactory. In contrast, our ghost point correction method significantly improves performance, achieving an F1 score of 91.5\%. Although this is 2.2\% lower than the F1 score in controlled experiments, the more complex real-world traffic conditions highlight the superior performance of our mmTunnel system.

\begin{figure}[t]
    \begin{minipage}[t]{0.48\linewidth}
        \centering
        \setlength{\abovecaptionskip}{0mm}
        \includegraphics[width=\linewidth]{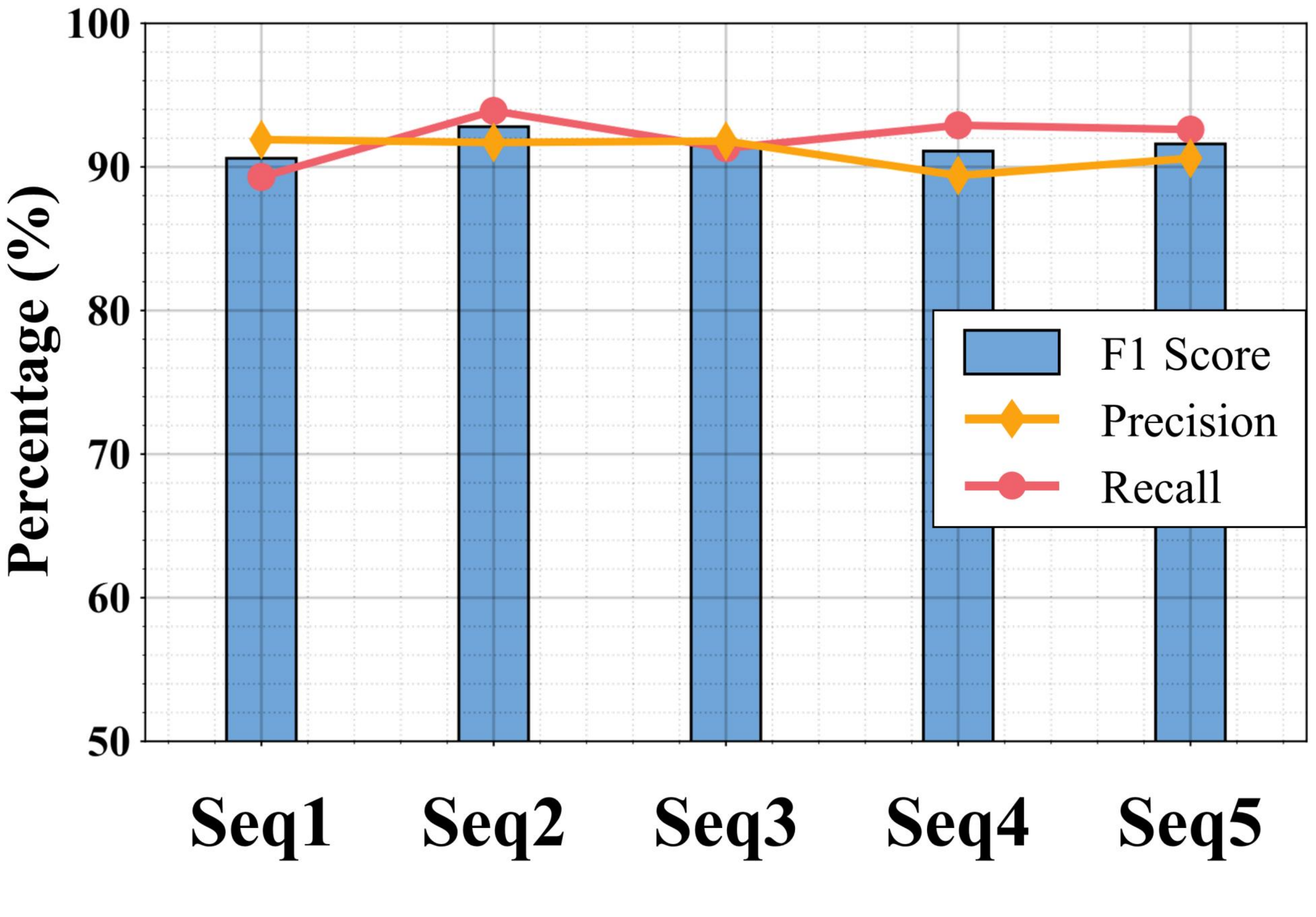}
        \caption{Vehicle detection performance across 5 sequences.
        }
        \label{fig:changan_seq}
    \end{minipage}
\hspace{1mm}
    \begin{minipage}[t]{0.48\linewidth}
        \centering
        \setlength{\abovecaptionskip}{0mm}
        \includegraphics[width=\linewidth]{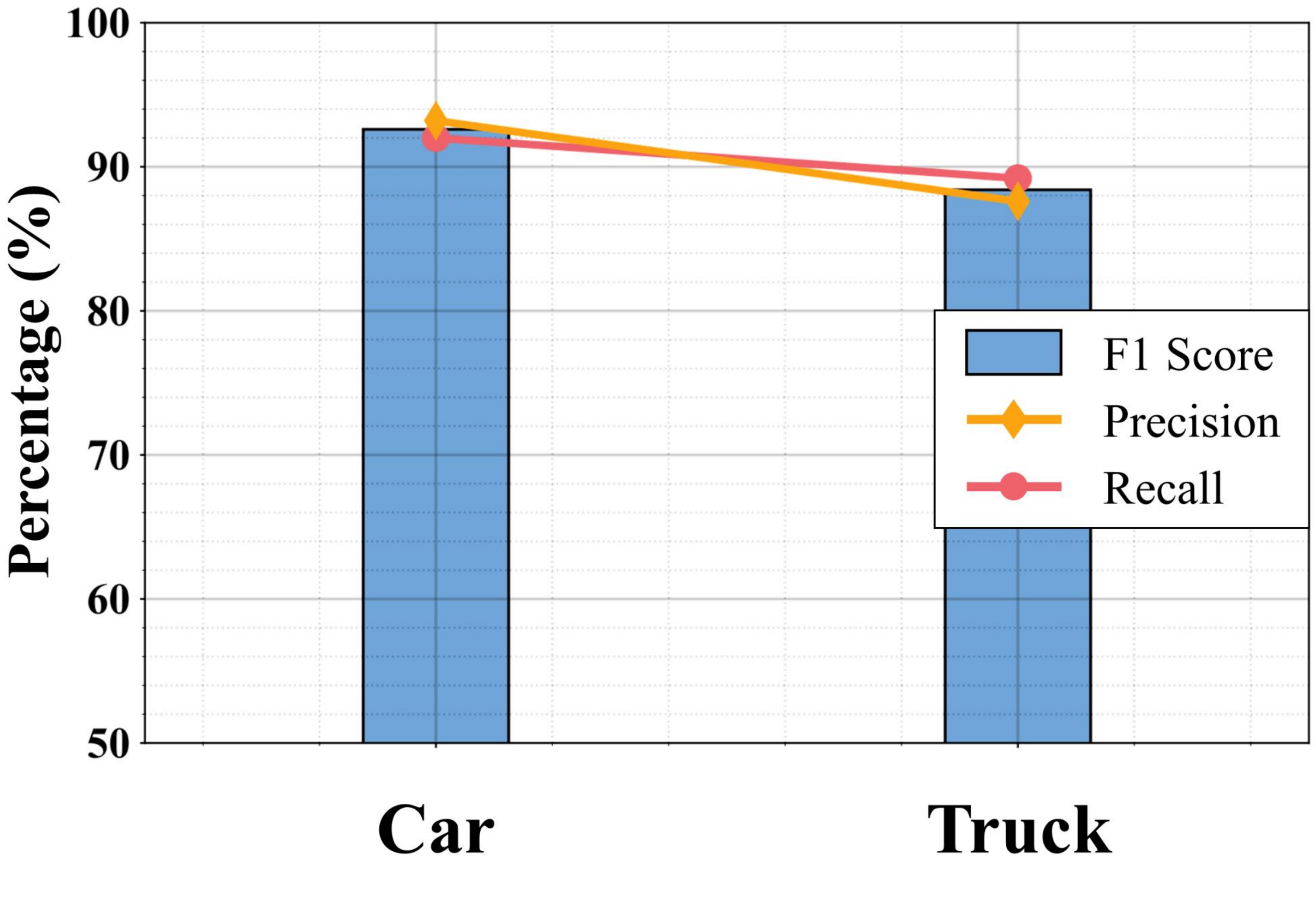}
        \caption{Vehicle detection performance across different vehicle types.
        }
        \label{fig:changan_type}
    \end{minipage}
\vspace{-3mm}
\end{figure}

We also evaluated the vehicle detection performance of mmTunnel across the five sequences in our dataset. As shown in Figure~\ref{fig:changan_seq}, the F1 score ranges from 90.6\% to 92.8\%. These results highlight the robustness of mmTunnel over extended periods in real-world scenarios.

\vspace{-1mm}
\subsubsection{Impact of Vehicle Type.}
Vehicle type significantly affects the distribution of radar points, thereby influencing detection performance. To evaluate this impact, we compared the detection performance for two main vehicle categories in our dataset: cars and trucks. As shown in Figure~\ref{fig:changan_type}, the detection performance for cars is notably higher, achieving an F1 score of 92.6\%. In contrast, the F1 score for trucks is 4.2\% lower than that for cars. This performance gap aligns with our controlled experiment results, where larger trucks were found to generate more ghost points, which are harder to correct and lead to more severe localization errors. We will further discuss the limitations of our system and potential solutions in Section~\ref{sec:limit}.

\vspace{-1mm}
\subsubsection{Occluded Vehicle Detection.}
\begin{figure}
    \centering
    \setlength{\abovecaptionskip}{0mm}
    \includegraphics[width=0.99\linewidth]{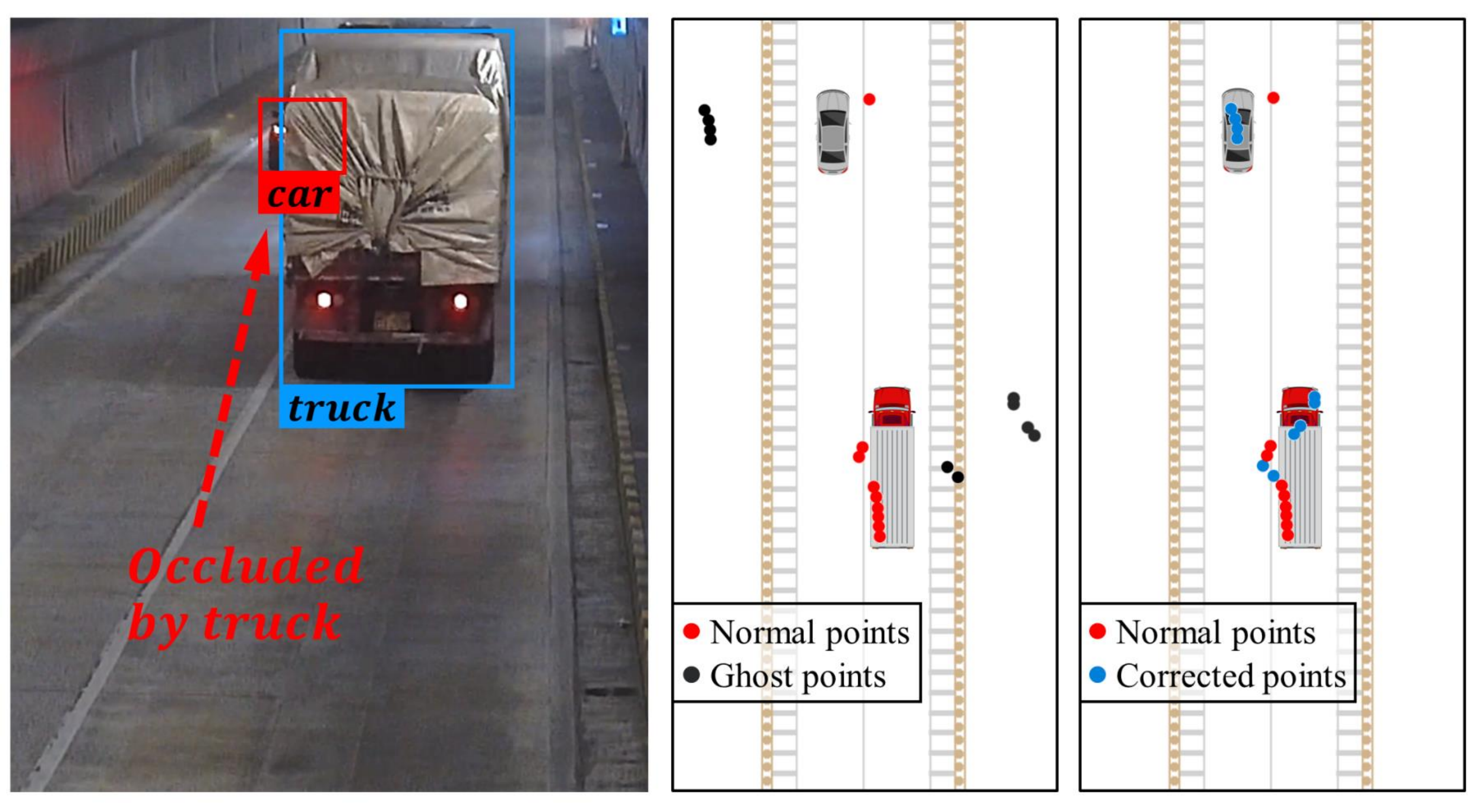}
    \caption{Visualization of occluded vehicle detection in the public tunnel. \textmd{Recall rate of the occluded car is improved from 22.8\% to 85.7\% by using the corrected points.}} 
     \label{fig:changan_vis}
\vspace{-4mm}
\end{figure}

In the public tunnel, due to the radar’s deployment location, the most common vehicle configuration during occlusion involves a truck driving in the right lane with a car obstructed ahead of it in the left lane. Figure~\ref{fig:changan_vis} illustrates an example where the car remains mostly occluded for an extended period. Our experiments reveal that using raw points results in a low recall rate of only 22.8\% for detecting this occluded car. However, by correcting the ghost points, our system significantly improves the recall rate to 85.7\%, demonstrating its effectiveness in detecting occluded vehicles under real-world traffic conditions.

\vspace{-2mm}
\section{DISCUSSION}\label{sec:limit}

\noindent \textbf{Potential Applications.} Looking beyond technical improvements, our system has several promising potential applications. Unlike traditional systems that use entry/exit counters or cameras, mmTunnel provides continuous, occlusion-resilient tracking of individual vehicles throughout a long tunnel. This capability is critical for real-time traffic monitoring. More importantly, it enables timely accident detection and emergency response by quickly identifying stopped vehicles or sudden traffic disruptions, thus enhancing both the safety and efficiency of tunnel operations.

\begin{figure}
    \centering
    \setlength{\abovecaptionskip}{0mm}
    \includegraphics[width=0.99\linewidth]{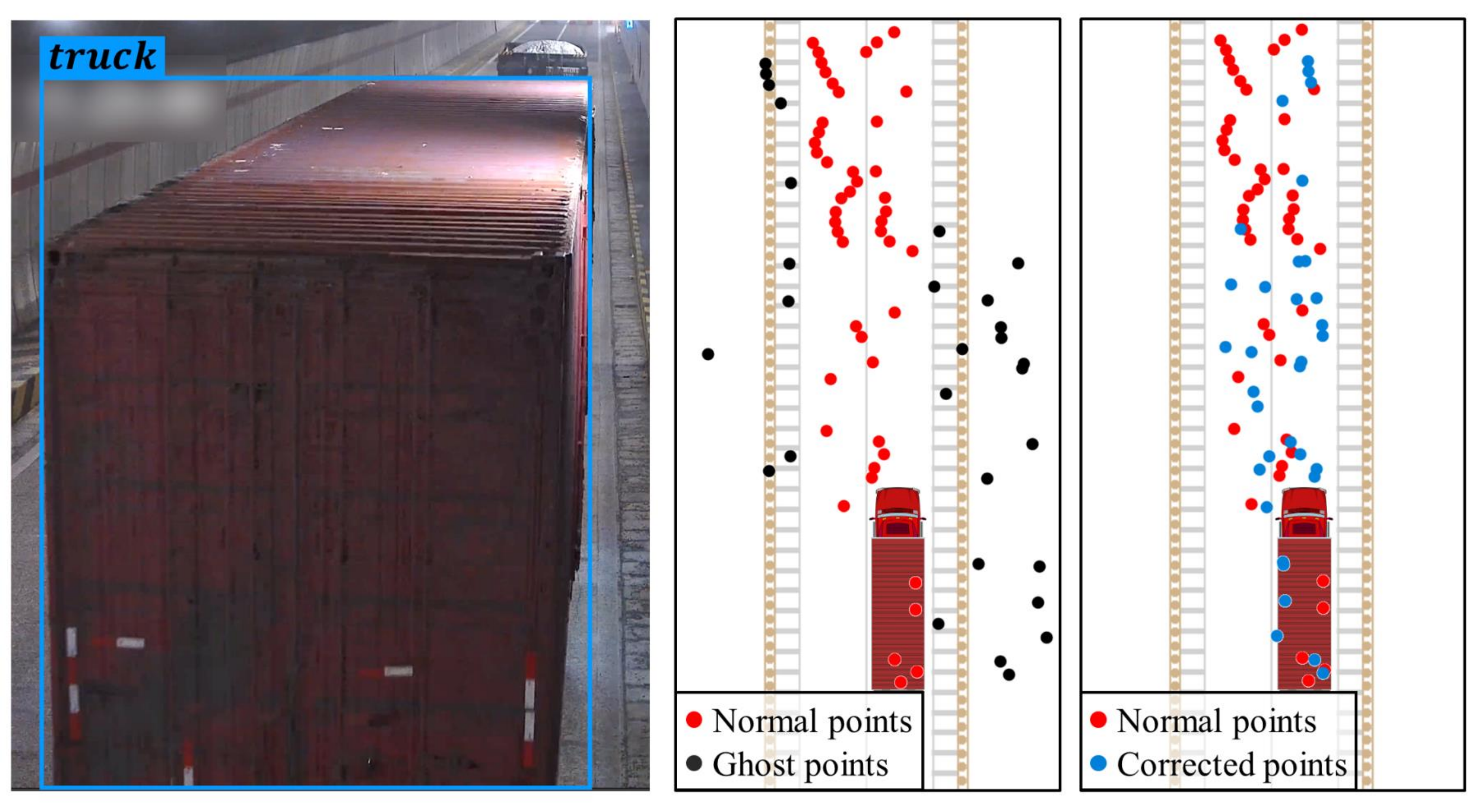}
    \caption{
    Visualization of radar points from a heavy-duty truck. \textmd{The truck’s large container causes complex signal reflections, leading to numerous ghost points.} 
    }
     \label{fig:changan_limit}
\vspace{-3mm}
\end{figure}

\noindent \textbf{System Limitations.} While mmTunnel has achieved high performance in many scenarios, it is important to acknowledge its limitations. 
First, we observed the occasional presence of heavy-duty box trucks in the public tunnel. Their large metal bodies cause the radar signals to reflect multiple times between  the vehicle and the tunnel surface. This generates a substantial number of ghost points that are scattered over a large area, often appearing both inside and outside the driving lanes. As shown in Figure~\ref{fig:changan_limit}, the system may incorrectly treat the ghost points inside the lanes as normal points and fail to correct those outside the lanes, leading to a decrease in detection performance for these trucks.
Second, the performance of our ghost point correction algorithm relies heavily on accurate radar calibration. In our experiments, this was achieved through careful, controlled measurements. However, such a labor-intensive process requires stopping all traffic for an extended time, which is not practical for real-world deployment. Calibration errors can lead to incorrect reflection path modeling, inaccurate ghost point correction, and ultimately, reduced detection accuracy.

\noindent \textbf{Future Work.} To address these limitations, we have several directions for future research. A key focus will be to develop a machine-learning-based classification algorithm to identify and eliminate ghost points within the driving lanes. Additionally, we plan to develop automatic calibration methods based on vehicle trajectories and known tunnel construction parameters, which would enable practical and efficient deployment without disrupting traffic.
We are also exploring the use of 4D mmWave radars, which can provide 3D coordinates and Doppler velocity of point clouds. This data could help infer vehicle types and dynamically adjust the assumed vehicle height ($H_{car}$) to improve ghost point correction. Furthermore, elevation information could be used to constrain the search space for reflection paths by considering ground plane consistency and the height distribution of the point clouds. This presents a promising avenue for developing a more robust ghost point correction method.

\section{RELATED WORK}
\textbf{Roadside Vehicle Detection.} Roadside perception is crucial for intelligent transportation systems. Previous works primarily use cameras or LiDARs for roadside vehicle detection based on deep learning algorithms~\cite{rope3d, Dair-v2x, bevdepth, bevheight, monogae, vips, vrf, trajmatch}. While these methods perform well on benchmark datasets, their sensing range is typically limited to 150 meters. Some researchers have proposed radar-camera fusion systems for far-range vehicle detection (over 300 meters)~\cite{farfusion, farfusion2} in open road environments. However, these systems do not account for the numerous ghost radar points present in tunnels, and they struggle with detecting occluded vehicles.

\vspace{1pt} \noindent \textbf{Ghost Point Identification.} The presence of ghost points is a critical challenge in mmWave sensing. Traditionally, these ghost points have been treated as interference, with researchers employing feature engineering techniques~\cite{ghost-rule, ghost-idf, mmOVD} or deep learning algorithms~\cite{Kraus-pointnet, Griebel-pointnet, rdgait} to identify and eliminate them. For instance, Wang et al.~\cite{rdgait} proposed a ghost detection algorithm based on velocity change patterns, achieving accurate ghost removal in complex indoor environments. However, in this work, we take a novel approach by utilizing the ghost points to enhance vehicle detection.

\vspace{1pt} \noindent \textbf{Radar-based NLOS Detection.} 
Many researchers have investigated non-line-of-sight (NLOS) detection using mmWave radar. For example, several works exploit wall-reflected signals to detect or image objects hidden behind corners in indoor environments~\cite{corner-radar, corner1, corner2, corner3, corner9, corner10, corner-mobicom}. While achieving high accuracy, the detection ranges in these studies are typically limited to about 10~m. Other researchers have explored NLOS detection in automotive scenarios~\cite{corner4, corner5, corner6, corner7, mmOVD, street-corner}, focusing on detecting occluded vehicles behind corners or other vehicles. However, these approaches generally consider only 2D reflections involving the ground, vertical walls, or other planar reflectors. The complex 3D reflections and far-range detection requirements in tunnels render such methods unsuitable for our scenario.




\vspace{-2mm}
\section{CONCLUSION}
We have presented mmTunnel, a practical roadside radar system designed for far-range vehicle detection in tunnels. Our system uniquely leverages ghost points---typically considered interference in previous works---to accurately recover vehicle positions. We are able to detect and localize vehicles in tunnels in real time, even in cases of complete occlusion. We believe that the design, implementation, and evaluation of our system make significant contributions to the field of intelligent transportation.
\vspace{-6pt}
\section{ACKNOWLEDGMENTS} \label{sec:ackn}
This work was supported by the National Natural Science Foundation of China (No. 62332016) and the Key Research Program of Frontier Sciences, CAS (No. ZDBS-LY-JSC001).

\bibliographystyle{ACM-Reference-Format}
\bibliography{bib/mmw}

\appendix

\newpage
\section{CURVED SURFACE MODELING}\label{apd-curve}
In Section~\ref{tunnel-modeling}, we present our segmentation-based modeling method. Here, we explain why curved surface modeling would lead to complex computation and make real-time processing infeasible.

When the tunnel is segmented into multiple planes, after calculating the distance from the ghost point to the true point ($dist_{G2T}$ in Equation~\ref{dist_G2T}), we can obtain the position of true point by moving the ghost point $dist_{G2T}$ in the direction perpendicular to the centerline. However, when the tunnel is modeled as a curved surface, the slope of centerline is not constant. This requires additional calculations to determine the position of the reflection point.

Specifically, we denote the position of radar as $(x_{rad}, y_{rad})$ and the position of ghost point as $(x_g, y_g)$. The reflection point, denoted as $R=(x_r, y_r)$, should lie on the line connecting the radar and the ghost point. This relationship can be described by the equation:
\begin{equation}\label{eq-18}
    \small y_r=b_1x_r + b_0,
\end{equation}
where $b_1 = \frac{y_g - y_{rad}}{x_g - x_{rad}}$, $b_0 = \frac{y_{rad}x_g - y_gx_{rad}}{x_g - x_{rad}}$. Assuming the distance from the reflection point to the centerline in the top-view coordinate system is $d$, with the nearest point on the centerline denoted as $S=(x_s, y_s)$, the line segment $\overrightarrow{SR}$ should be perpendicular to the tangent of the centerline at point $S$.
\begin{equation}\label{eq-19}
    \small y_r - y_s = -\frac{1}{m_s}\times (x_r - x_s),
\end{equation}
where $m_s = 3a_3x_s^2 + 2a_2x_s + a_1$, representing the slope of the tangent of the centerline at point $S$. By combining Equation~\ref{eq-18} and \ref{eq-19}, we have:
\begin{equation}
    \small x_r = \frac{x_s + m_sy_s-m_sb_0}{b_1m_s + 1}.
\end{equation}

Further, since $\left| \overrightarrow{SR} \right| = d$:
\vspace{-2mm}
\begin{equation}
    \small \left| \overrightarrow{SR} \right| = \left| x_r - x_s \right| \times \sqrt{\frac{1}{m_s^2} + 1} = d
\end{equation}
\begin{equation}\label{eq-22}
    \small \Rightarrow \frac{a_3x_s^3 + a_2x_s^2 + a_1x_s + a_0 - b_1x_s - b_0}{b_1\times(3a_3x_s^2 + 2a_2x_s + a_1) + 1} = \frac{d}{\sqrt{(3a_3x_s^2 + 2a_2x_s + a_1)^2 + 1}}.
\end{equation}

By solving Equation~\ref{eq-22}, we can obtain the slope and intercept of the tangent of the centerline at point $S$:
\begin{equation}
\left\{
    \begin{aligned}
        \small m_s &= 3a_3x_s^2 + 2a_2x_s + a_1 \\
        \small b_s &= a_3x_s^3 + a_2x_s^2 + a_1x_s + a_0 - m_sx_s
    \end{aligned}
    \right.
\end{equation}
Here, $m_s$ and $b_s$ are analogous to $m$ and $b$ in Equation~\ref{eq-Center}, but specifically calculated for the curved surface model. The remaining steps to correct ghost points are the same as those outlined in Section~\ref{sec:ghost-correct} and will not be repeated here.

We have implemented this method and compared it with our segmentation-based modeling method, detailed in Section~\ref{sec:evaluation}. Notably, Equation~\ref{eq-22} is not analytically solvable, so we use Newton's method to compute its numerical solution. This introduces significant delays, making real-time processing infeasible.

\section{DERIVATION OF EQUATION (4)}\label{apd-deriv}
\subsection{Error of Cross-section Segmentation}
\vspace{-3mm}
\begin{figure}[h]
    \centering
    \setlength{\abovecaptionskip}{0mm}
    \includegraphics[width=0.98\linewidth]{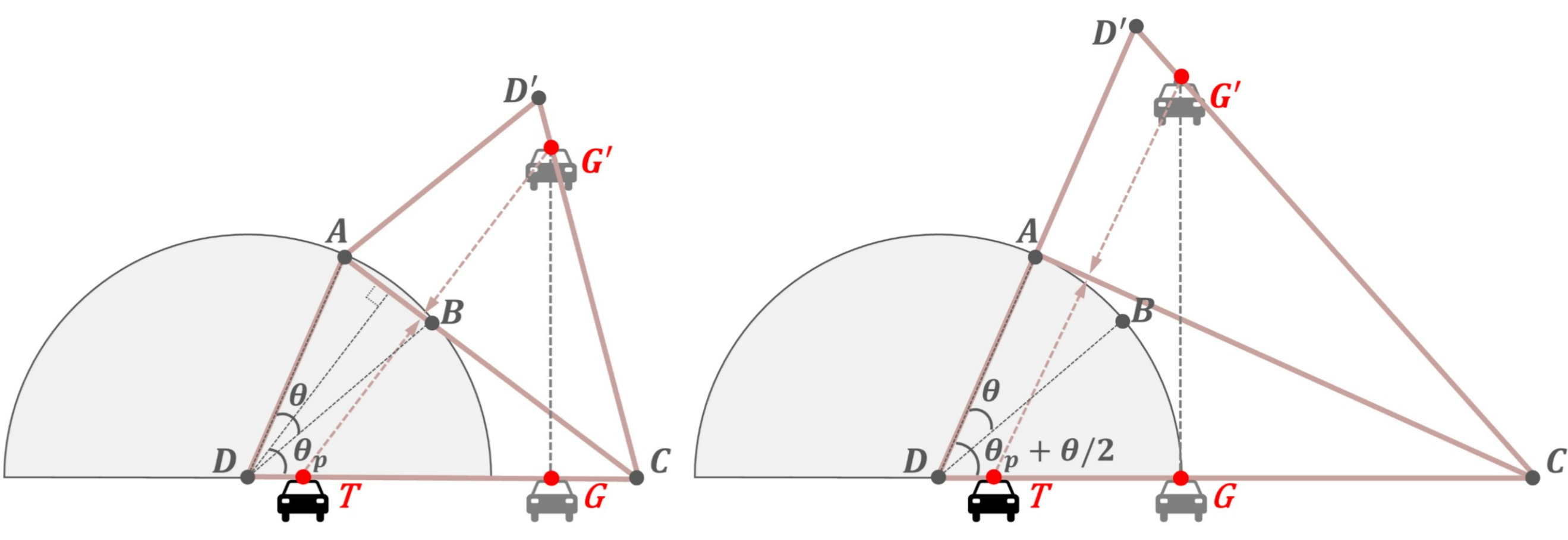}
    \caption{Reflections at two ends of the segment cause the maximum errors.} 
     \label{fig:appendix1}
\vspace{-2mm}
\end{figure}

As shown in Figure~\ref{fig:appendix1}, cross-section segmentation can introduce localization errors. To facilitate calculations, we assume $H_{center}=H_{car}$, as their values are nearly identical in our test tunnels. The center of the circular cross-section is denoted as $A$, and the true vehicle position is denoted as $T$. Assuming the reflection occurs on the segment $AB$, the position of the ghost point $G$ can be derived, as illustrated in the left sub-figure in Figure~\ref{fig:appendix1}:
\begin{equation}
\left\{
    \begin{aligned}
        \small |CD| &= \frac{R_{tunnel}\cos{\frac{\theta}{2}}}{\cos{\theta_p}} \\
        \small |CG| &= (|CD| - |DT|) \times \cos{2\theta_p}
    \end{aligned}
    \right.
\end{equation}
Further, we have $|DG| = |CD| + |CG|$:
\begin{equation}\label{eq-25}
    \small |DG| = |DT| + 2R_{tunnel} \cos{\frac{\theta}{2}} \cos{\theta_p} - 2|DT| \cos^2{\theta_p},
\end{equation}
where $\theta$ is the sector angle of each segment, and $\theta_p$ is the angle between the perpendicular bisector of $AB$ and the ground plane. 

However, the above derivations approximate $AB$ as a straight line segment, which introduces small errors because the slope of the tangent at the reflection point is not constant. Clearly, the error reaches its maximum when the reflection occurs at the two ends of the segment. The right sub-figure in Figure~\ref{fig:appendix1} shows an example. Similarly, we can calculate the position of ghost point $G$ in such cases:
\begin{equation}\label{eq-26}
    \small |DG|_{end} = |DT| + 2R_{tunnel}\cos{\left( \theta_p \pm \frac{\theta}{2} \right)} - 2|DT|\cos^2{\left( \theta_p \pm \frac{\theta}{2} \right)}.
\end{equation}

By combining Equation~\ref{eq-25} and \ref{eq-26}, we obtain the error function:
\begin{equation}
\begin{aligned}
    \small E_c =& \Big| |DG| - |DG|_{end} \Big| \\
    \small =& \left| 2|DT|\left( \cos^2\left(\theta_p \pm \frac{\theta}{2}\right) - \cos^2{\theta_p}\right) \pm 2R_{tunnel}\sin{\theta_p}\sin{\frac{\theta}{2}} \right|.
\end{aligned}
\end{equation}
When $\theta_p = \frac{\pi}{2}$, the value of $E_c$ reaches the maximum:
\begin{equation}
\begin{aligned}
    \small E_c &\leq 2R_{tunnel}\sin{\frac{\theta}{2}} + 2|DT|\sin^2\frac{\theta}{2} \\
    \small \Rightarrow E_c &< 2R_{tunnel}\left( \sin{\frac{\theta}{2}} + \sin^2\frac{\theta}{2} \right).
\end{aligned}
\end{equation}

\subsection{Error of Tunnel-path Segmentation}
\vspace{-3mm}
\begin{figure}[h]
    \centering
    \setlength{\abovecaptionskip}{0mm}
    \includegraphics[width=0.5\linewidth]{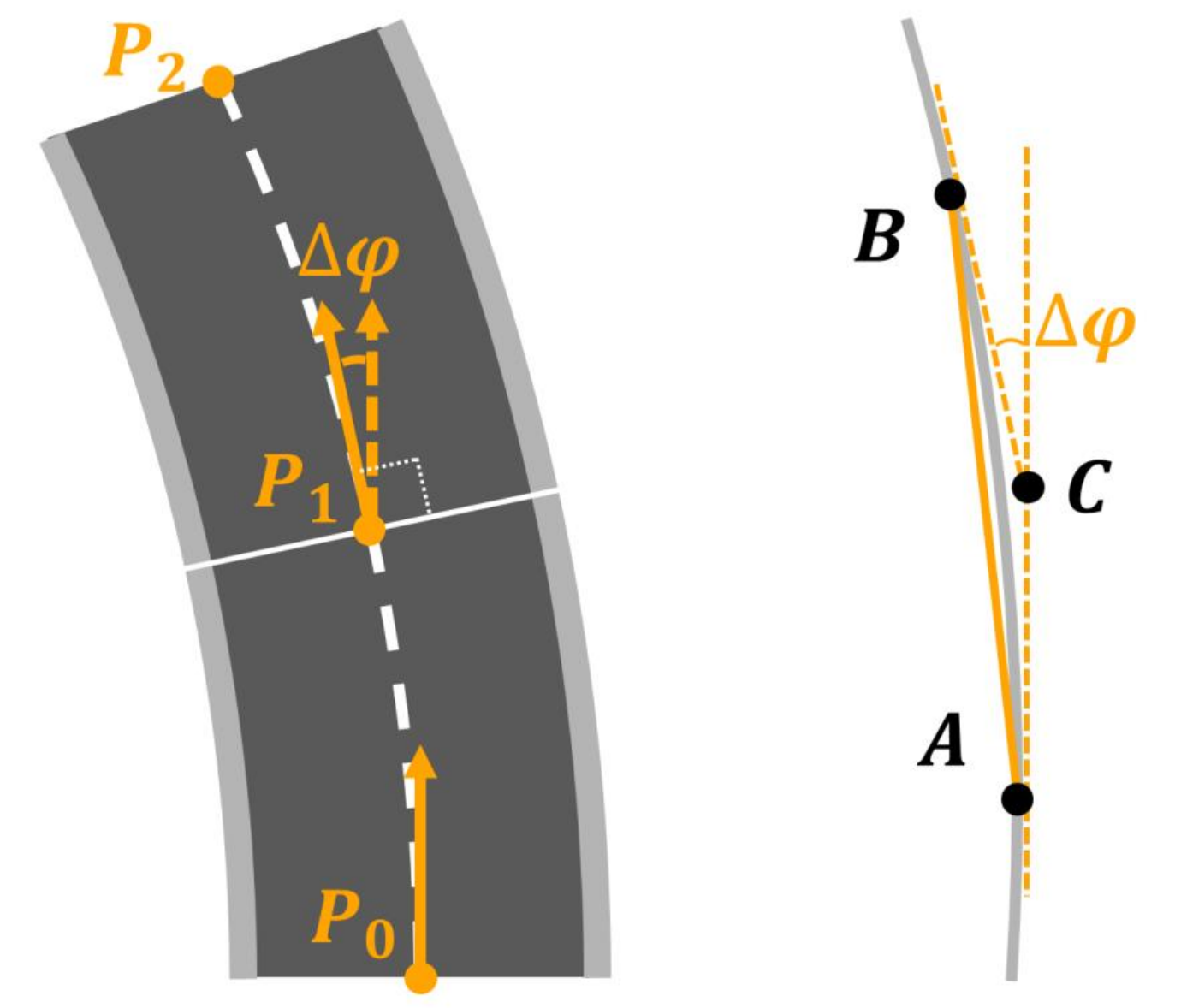}
    \caption{Approximation errors from the tunnel-path segmentation.} 
     \label{fig:appendix2}
\vspace{-3mm}
\end{figure}
As shown in Figure~\ref{fig:appendix2}, tunnel-path segmentation can also introduce errors. For ease of understanding, let's focus on the centerline of the tunnel path. We denote the two ends of a segment as $A$ and $B$. The tangents of the centerline at $A$ and $B$ have an intersection point, denoted as $C$. 

Therefore, the error between the original centerline and the line segment $AB$ is bounded by the distance between $C$ and line $AB$:
\begin{equation}
\begin{aligned}
    \small E_p &\leq |AC| \sin{\angle BAC} \\
    &< |AC| \sin\Delta\varphi \\
    &< |AB| \tan\Delta\varphi
\end{aligned}
\end{equation}

It can be seen that the error is influenced by both the length of the path segment and the threshold angle $\Delta\varphi$. Therefore, we constrain the length of each path segment with a maximum threshold $L_{max}$:
\begin{equation}
    \small E_p < L_{max}\tan\Delta\varphi.
\end{equation}

\end{document}